\documentstyle[12pt]{article}
\def\journal{\topmargin .3in	\oddsidemargin .5in
	\headheight 0pt	\headsep 0pt
	\textwidth 5.625in 
	\textheight 8.25in 
	\marginparwidth 1.5in
	\parindent 2em
	\parskip .5ex plus .1ex		\jot = 1.5ex}
\journal

\catcode`\@=11
\def\marginnote#1{}
\newcount\hour
\newcount\minute
\newtoks\amorpm
\hour=\time\divide\hour by60
\minute=\time{\multiply\hour by60 \global\advance\minute by-\hour}
\edef\standardtime{{\ifnum\hour<12 \global\amorpm={am}%
	\else\global\amorpm={pm}\advance\hour by-12 \fi
	\ifnum\hour=0 \hour=12 \fi
	\number\hour:\ifnum\minute<10 0\fi\number\minute\the\amorpm}}
\edef\militarytime{\number\hour:\ifnum\minute<10 0\fi\number\minute}
\def\draftlabel#1{{\@bsphack\if@filesw {\let\thepage\relax
   \xdef\@gtempa{\write\@auxout{\string
      \newlabel{#1}{{\@currentlabel}{\thepage}}}}}\@gtempa
   \if@nobreak \ifvmode\nobreak\fi\fi\fi\@esphack}
	\gdef\@eqnlabel{#1}}
\def\@eqnlabel{}
\def\@vacuum{}
\def\draftmarginnote#1{\marginpar{\raggedright\scriptsize\tt#1}}
\def\draft{\oddsidemargin -.5truein
	\def\@oddfoot{\sl preliminary draft \hfil 
	\rm\thepage\hfil\sl\today\quad\militarytime}
	\let\@evenfoot\@oddfoot	\overfullrule 3pt
	\let\label=\draftlabel 
	\let\marginnote=\draftmarginnote
   \def\@eqnnum{(\theequation)\rlap{\kern\marginparsep\tt\@eqnlabel}%
\global\let\@eqnlabel\@vacuum}  }
\def\preprint{\twocolumn\sloppy\flushbottom\parindent 2em
	\leftmargini 2em\leftmarginv .5em\leftmarginvi .5em
	\oddsidemargin -.5in	\evensidemargin -.5in
	\columnsep .4in	\footheight 0pt
	\textwidth 10in	\topmargin  -.4in
	\headheight 12pt \topskip .4in
	\textheight 7.1in \footskip 0pt 
	\def\@oddhead{\thepage\hfil\addtocounter{page}{1}\thepage}
	\let\@evenhead\@oddhead	\def\@oddfoot{}	\def\@evenfoot{} }
\def\numberbysection{\@addtoreset{equation}{section}
	\def\theequation{\thesection.\arabic{equation}}}
\def\underline#1{\relax\ifmmode\@@underline#1\else
	$\@@underline{\hbox{#1}}$\relax\fi}
\def\titlepage{\@restonecolfalse\if@twocolumn\@restonecoltrue\onecolumn
     \else \newpage \fi \thispagestyle{empty}\c@page\z@	
	\def\thefootnote{\fnsymbol{footnote}} }
\def\endtitlepage{\if@restonecol\twocolumn \else \newpage \fi
	\def\thefootnote{\arabic{footnote}} 
	\setcounter{footnote}{0}}  
\catcode`@=12
\relax
\def\figcap{\section*{Figure Captions\markboth
	{FIGURECAPTIONS}{FIGURECAPTIONS}}\list
	{Figure \arabic{enumi}:\hfill}{\settowidth\labelwidth{Figure 999:}
	\leftmargin\labelwidth 
	\advance\leftmargin\labelsep\usecounter{enumi}}}
 \relax
\def\tablecap{\section*{Table Captions\markboth
	{TABLECAPTIONS}{TABLECAPTIONS}}\list
	{Table \arabic{enumi}:\hfill}{\settowidth\labelwidth{Table 999:}
	\leftmargin\labelwidth 
	\advance\leftmargin\labelsep\usecounter{enumi}}}
 \relax
\def\reflist{\section*{References\markboth
	{REFLIST}{REFLIST}}\list
	{[\arabic{enumi}]\hfill}{\settowidth\labelwidth{[999]}
	\leftmargin\labelwidth 
	\advance\leftmargin\labelsep\usecounter{enumi}}}
 \relax
\makeatletter
\newcounter{pubctr}
\def\publist{\@ifnextchar[{\@publist}{\@@publist}}
\def\@publist[#1]{\list
	{[\arabic{pubctr}]\hfill}{\settowidth\labelwidth{[999]}
	\leftmargin\labelwidth 
	\advance\leftmargin\labelsep
	\@nmbrlisttrue\def\@listctr{pubctr}
	\setcounter{pubctr}{#1}\addtocounter{pubctr}{-1}}}
\def\@@publist{\list
	{[\arabic{pubctr}]\hfill}{\settowidth\labelwidth{[999]}
	\leftmargin\labelwidth 
	\advance\leftmargin\labelsep
	\@nmbrlisttrue\def\@listctr{pubctr}}}
 \relax
\makeatother
\catcode`\@=11
\def\section{\@startsection {section}{1}{0pt}{-3.5ex plus -1ex minus 
 -.2ex}{2.3ex plus .2ex}{\raggedright\large\bf}}
\catcode`\@=12
\newskip\humongous \humongous=0pt plus 1000pt minus 1000pt
\def\caja{\mathsurround=0pt}

\newif\ifdtup
\def\panorama{\global\dtuptrue \openup1\jot \caja
	\everycr{\noalign{\ifdtup \global\dtupfalse
	\vskip-\lineskiplimit \vskip\normallineskiplimit
	\else \penalty\interdisplaylinepenalty \fi}}}
\def\eqalignno#1{\panorama \tabskip=\humongous
	\halign to\displaywidth{\hfil$\displaystyle{##}$
	\tabskip=0pt&$\displaystyle{{}##}$\hfil
	\tabskip=\humongous&\llap{$##$}\tabskip=0pt
	\crcr#1\crcr}}

\def\oldreffmt#1{\rlap{[#1]} \hbox to 2\parindent{}}

\def\figfmt#1{\rlap{Figure {#1}} \hbox to 1in{}}

\def\Im{\mathop{\rm Im}}
\def\Re{\mathop{\rm Re}}

\def\VEV#1{\left\langle #1\right\rangle}
\let\vev\VEV

\def\half{{1\over 2}}
\def\beq{\begin{equation}}
\def\eeq{\end{equation}}

\def\bea{\begin{eqnarray}}                                 

\def\eea{\end{eqnarray}}
\def\np#1#2#3{           {\it Nucl. Phys. }{\bf #1}, #2 (19#3)}
\def\pl#1#2#3{           {\it Phys. Lett. }{\bf #1}, #2 (19#3)}
\def\pr#1#2#3{           {\it Phys. Rev. }{\bf #1}, #2 (19#3)}

\def\prl#1#2#3{          {\it Phys. Rev. Lett. }{\bf #1}, #2 (19#3)}

\def\zp#1#2#3{           {\it Zeit. fur Physik }{\bf #1}, #2 (19#3)}
\catcode`\@=11
\def\eqnarray{\stepcounter{equation}\let\@currentlabel=\theequation
\global\@eqnswtrue
\global\@eqcnt\z@\tabskip\@centering\let\\=\@eqncr
\gdef\@@fix{}\def\eqno##1{\gdef\@@fix{##1}}%
$$\halign to \displaywidth\bgroup\@eqnsel\hskip\@centering
  $\displaystyle\tabskip\z@{##}$&\global\@eqcnt\@ne
  \hskip 2\arraycolsep \hfil${##}$\hfil
  &\global\@eqcnt\tw@ \hskip 2\arraycolsep $\displaystyle\tabskip\z@{##}$\hfil
   \tabskip\@centering&\llap{##}\tabskip\z@\cr}

\def\@@eqncr{\let\@tempa\relax
    \ifcase\@eqcnt \def\@tempa{& & &}\or \def\@tempa{& &}
      \else \def\@tempa{&}\fi
     \@tempa \if@eqnsw\@eqnnum\stepcounter{equation}\else\@@fix\gdef\@@fix{}\fi
     \global\@eqnswtrue\global\@eqcnt\z@\cr}

\catcode`\@=12
\font\tenbifull=cmmib10 
\font\tenbimed=cmmib10 scaled 800
\font\tenbismall=cmmib10 scaled 666
\relax

\def\wul{{W_{U_L}}}
\def\wur{{W_{U_R}}}
\def\wdl{{W_{D_L}}}
\def\wdr{{W_{D_R}}}
\def\wel{{W_{E_L}}}
\def\wer{{W_{E_R}}}
\def\wulr{{W_{U_{L(R)}}}}
\def\wurl{{W_{U_{R(L)}}}}
\def\wdlr{{W_{D_{L(R)}}}}
\def\wdrl{{W_{D_{R(L)}}}}
\def\welr{{W_{E_{L(R)}}}}
\def\werl{{W_{E_{R(L)}}}}
\def\gsim{\,\raisebox{-.5ex}{\rlap{$\sim$}}\raisebox{.5ex}{$>$}\,}
\def\lsim{\,\raisebox{-.5ex}{\rlap{$\sim$}}\raisebox{.5ex}{$<$}\,}
\def\ba{\begin{eqnarray*}}
\def\ea{\end{eqnarray*}}
\def\vub{\mbox{$V_{ub}$}}
\def\vcb{\mbox{$V_{cb}$}}
\def\vtd{\mbox{$V_{td}$}}
\def\vts{\mbox{$V_{ts}$}}

\def\vus{\mbox{$V_{us}$}}
\def\sc{\mbox{$\sin \theta_c$}}
\def\tb{\mbox{$\tan \beta$}}
\def\wt{\widetilde}
\def\gev{\mbox{GeV}}
\def\mev{\mbox{MeV}}

\begin{document}
\begin{titlepage}
\begin{center}
\today     \hfill    LBL-37894 \\
          \hfill    UCB-PTH-95/36 \\
\hfill hep-ph/9601262 \\

\vskip .5in

{\large \bf A Supersymmetric Theory of Flavor with\\ 
   Radiative Fermion Masses}
\footnote{This work was supported in part by the Director, Office of
Energy Research, Office of High Energy and Nuclear Physics, Division of
High Energy Physics of the U.S. Department of Energy under Contract
DE-AC03-76SF00098 and in part by the National Science Foundation under
grant PHY-90-21139.}

\vskip .5in
{\bf Nima Arkani-Hamed\\
   Hsin-Chia Cheng}\\
and\\
{\bf L.J. Hall}\\
\vskip .20in

{\em Theoretical Physics Group\\
    Lawrence Berkeley National Laboratory\\
and\\
Department of Physics\\
      University of California\\
    Berkeley, California 94720}
\end{center}

\vskip .5in

\begin{abstract}
Supersymmetric theories involving a spontaneously broken flavor 
symmetry can lead to fermion masses which vanish at tree level 
but are generated by radiative corrections. In the context of supersymmetric
theories with minimal low energy field content we discuss which fermion
masses and mixings may be obtained radiatively, and find that 
constraints from flavor changing phenomenology imply that only the 
first generation fermion masses and some (but not all) CKM mixings can 
naturally come from radiative corrections. We also consider general 
conditions on theories of flavor which guarantee the existence of tree 
level massless fermions while having non-trivial CKM matrix elements at 
tree level. Two complete models of flavor are presented. In the first 
model, all first generation fermion masses are radiatively generated. 
In the second model, the electron and up quark mass are due to 
radiative corrections whereas the down mass appears at tree level, 
as does a successful prediction for the Cabibbo angle 
 $\sin \theta_c = \sqrt{m_d/m_s}$.
This model can be embedded in the flipped $SU(5)$ grand unified theory. 
\end{abstract}
\end{titlepage}
\renewcommand{\thepage}{\roman{page}}
\setcounter{page}{2}
\mbox{ }

\vskip 1in

\begin{center}
{\bf Disclaimer}
\end{center}

\vskip .2in

\begin{scriptsize}
\begin{quotation}
This document was prepared as an account of work sponsored by the United
States Government. While this document is believed to contain correct
 information, neither the United States Government nor any agency
thereof, nor The Regents of the University of California, nor any of their
employees, makes any warranty, express or implied, or assumes any legal
liability or responsibility for the accuracy, completeness, or usefulness
of any information, apparatus, product, or process disclosed, or represents
that its use would not infringe privately owned rights.  Reference herein
to any specific commercial products process, or service by its trade name,
trademark, manufacturer, or otherwise, does not necessarily constitute or
imply its endorsement, recommendation, or favoring by the United States
Government or any agency thereof, or The Regents of the University of
California.  The views and opinions of authors expressed herein do not
necessarily state or reflect those of the United States Government or any
agency thereof of The Regents of the University of California and shall
not be used for advertising or product endorsement purposes.
\end{quotation}
\end{scriptsize}

\vskip 2in

\begin{center}
\begin{small}
{\it Lawrence Berkeley Laboratory is an equal opportunity employer.}
\end{small}
\end{center}

\newpage
\renewcommand{\thepage}{\arabic{page}}
\setcounter{page}{1}

\section{Introduction}

\hspace{\parindent} A complete supersymmetric theory of flavor must address
both the fermion mass problem and the flavor changing problem \cite{DG}. An
early proposal to address the flavor changing problem by invoking a $U(N)$
flavor symmetry of the Kahler potential in supergravity \cite{HLW} was very
incomplete; it did not address how the symmetry could be broken to get the
fermion mass interactions of the superpotential. By studying the spontaneous
breaking of flavor symmetries, one can study both issues simultaneously
\cite{DLK}, opening the door to a new field of flavor model building. Although
there is considerable freedom in the choice of the flavor symmetry group and
the pattern of symmetry breaking, the enterprise is nevertheless constrained by
the direct link between the flavor changing and fermion mass problems. Many
candidate theories of fermion masses are excluded by flavor changing
phenomenology. In this paper we study the possibility that
some fermion masses arise radiatively, which requires large flavor
changing interactions of the squarks or sleptons. Hence theories
of flavor, based on spontaneously broken flavor symmetries, which involve 
radiative fermion masses, are very highly constrained by flavor changing
phenomenology.

Flavor symmetries should forbid Yukawa
couplings of the light fermions. After the flavor symmetries are
broken, the light generation fermions should acquire small Yukawa couplings. 
Many
models of fermion masses use the Froggatt-Nielsen mechanism \cite{FN} to 
generate small Yukawa
couplings: assuming a flavor symmetry is broken by the VEV of some fields 
$\VEV{\phi}$, and after integrating out heavy states of mass $M$,
 one can get light
generation Yukawa couplings suppressed by ${\VEV{\phi}\over M}$.
 This mechanism can
naturally generate second generation Yukawa couplings, but in order to ensure
small enough first generation Yukawa couplings one usually has to assume
contrived representations of the flavor group and/or contrived patterns of
flavor breaking. There is, however, another possibility for generating small
Yukawa couplings: if generated radiatively, they are suppressed by the loop
factor ${1\over 16 \pi^2}$.  This intriguing possibility has been extensively
studied in the literature\cite{Ma}. 
A universal feature of all models must be that
an ``accidental" chiral symmetry is present in the Yukawa sector to force a zero
Yukawa coupling at tree level, while this symmetry must be  broken in
another sector of the theory in order for the Yukawa coupling to be radiatively
generated. As we pointed out in \cite {ACH1}, supersymmetric theories can
provide a natural way for this to happen: the constraints of holomorphy can
force the superpotential to have accidental symmetries not shared by the
$D$-terms. Given that the supersymmetric extension of the standard model is
of interest for other reasons, we are naturally led to explore the idea of
radiative fermion masses in supersymmetric models. To be specific, we consider
supersymmetric $SU(3) \times SU(2) \times U(1)$ theories with minimal 
low energy field
content, i.e.\ we do not consider extra Higgses or extra families etc. We will
find that, with this assumption, the set of possibilities for radiative fermion
masses is highly constrained, and yields robust experimental predictions.

The outline of this paper is as follows. In section 2 we consider general
possibilities for radiative fermion masses in 
supersymmetric theories with minimal low-energy field
content, and conclude that, quite generally, only the lightest generation can
be obtained radiatively. In section 3 we discuss phenomenological constraints
and consequences which follow from 
 generating the lightest generation radiatively. In the
subsequent sections, we consider issues related to building models which
naturally implement radiative fermion Yukawa couplings for the first generation:
In section 4, we discuss some general properties such models should have; 
and in
section 5 we extend the lepton model presented in \cite{ACH1} to the quark
sector. Our conclusions are drawn in section 6.

\section{General possibilities for radiative fermion masses}

\hspace{\parindent} We now consider 
general possibilities for radiatively generated Yukawa
couplings in supersymmetric theories with minimal low energy field content. We
know that, in the limit of exact supersymmetry, a Yukawa coupling which is zero
at tree level will never be generated radiatively. Thus, in order to have
radiative Yukawa couplings, we need soft supersymetry breaking operators which,
further, must explicitly break the chiral symmetries associated with the zero
Yukawa couplings of the superpotential. Also, the particles in the radiative
loop must be at the weak scale: since the generated Yukawa coupling $\lambda$ is
dimensionless and vanishes in the limit $m_S$ (the supersymmetry breaking scale)
goes to zero, 
we must have $\lambda \sim {{1\over 16 \pi^2}{ m_S\over M}}$, where
$M$ is a typical mass for the particles in the loop. Thus, $M$ must be near the
weak scale (rather than the GUT or Planck scale) in order to generate large
enough Yukawa couplings. 

Thus, we   see that the  breaking of the flavor symmetries associated
with the zero Yukawa couplings must lie in the weak scale soft supersymmetry
breaking operators: the trilinear scalar $A$ terms and the soft scalar masses.
In this paper  we make the plausible assumptions that 
the flavor symmetry is not an $R$ symmetry and that
supersymmetry breaking fields are flavor singlets.
Then, the $A$ terms must respect the same
flavor symmetries as the the Yukawa couplings, since any flavor symmetry
forbidding $\int d^2 \theta f(\phi)$ (where $f(\phi)$ is some function of the
superfields $\phi$ in the theory) will also forbid $\int d^2 \theta \theta^2
f(\phi)$. Hence, all the flavor symmetry breaking responsible for
generating radiative fermion masses  resides in the scalar mass
matrices. (However, in appendix A, we repeat the analysis without this
assumption.  Requiring  our vacuum
to be  the global minimum of the potential and using  constraints from
flavor-changing neutral currents (FCNC), the $A$ terms are such that 
the conclusions of this section are
not greatly altered.)

For simplicity, let us work in the lepton sector, and consider the possibility
of radiatively generating $K$ lepton masses for $K =$ 3,2,1 in turn.

{\boldmath $K$}{\bf =3.} In this case, we have a vanishing tree level Yukawa 
matrix which
has a large $U(3)_\ell \times U(3)_e$ symmetry. By our assumption
that the flavor symmetry is not an $R$ symmetry and
that supersymmetry breaking fields do not carry flavor, the $A$ terms
must also vanish. But then, all the soft scalar mass matrices can be
simultaneously diagonalized, leaving an independent, unbroken $U(1)$ symmetry
acting on every superfield, preventing the radiative generation of any Yukawa
couplings.

{\boldmath $K$}{\bf =2.} Here, we only have the third generation Yukawa 
coupling at tree
level. This case is more interesting. We shall find that, although it is
possible to generate two Yukawa eigenvalues radiatively, strong constraints from
FCNC force the ratio of the (radiatively generated) first to second generation
Yukawa couplings to a value too small to be compatible with experiment.

Let us work in a basis where the Yukawa matrix {\boldmath $\lambda_E$} is
diagonal,
$$
\mbox{\boldmath $\lambda_E$} = \left(\begin{array}{lll} 0& 0& 0\\ 0 & 0 & 0
\\0 & 0 & 
\lambda \end{array}
\right). \eqno(2.1)    
$$
Since {\boldmath$\lambda_E$} is invariant under independent rotations of
the first two generation
left and right handed lepton superfields, we can make these rotations on the
left and right handed scalar masses ${\bf m}^2_{L(R)}$,
$$
{\bf m}^2_{L(R)} \rightarrow U_{L(R)} {\bf m}^2_{L(R)} U_{L(R)} ^\dagger ,
\eqno(2.2)
$$
where the $U_{L(R)}$ are unitary rotations in the upper 2 $\times$ 2 block,
$$
U_{L(R)} = \left(\begin{array}{lll} u_{L(R)} & 0\\ 0 & 1\end{array}\right).
\eqno(2.3)
$$
If we write 
$$
{\bf m}^2_L = \left(\begin{array}{lll} m^2_{2 \times 2} & m^{2}_{2 \times 1}
\\ m^{2 \dagger}_{2 \times 1} & m^2_{33}
\end{array}\right),
\eqno(2.4)
$$
then under $U_L$ we have
$$
{\bf m}^2_L \rightarrow \left(\begin{array}{lll} u_L m^2_{2 \times 2} 
u_L^\dagger & u_L m^2_{2 \times 1}\\
{u_L}^\dagger m^{2 \dagger}_{2 \times
1} & m^2_{33}\end{array}\right),
\eqno(2.5)
$$
and we can choose $u_L$ so that 
$$
u_L m^2_{2 \times 1} = \left(\begin{array}{l} 0 \\ m^2_{23}\end{array}\right).
\eqno(2.6)
$$
Thus, we can choose a basis where
 the 1-3 and 3-1 entries of ${\bf m}^2_L$ are 0,
and similarly for ${\bf m}^2_R$; the scalar masses
have the form
$$
{\bf m}^2_{L(R)} =
 \left(\begin{array}{lll} m^2_1 & \delta m^2_{12} & 0\\ 
\delta m^{2 *}_{12} & m^2_2 & \Delta m^2_{23}\\
0 & \Delta m^{2 *}_{23} & m^2_3\end{array}\right)_{L(R)}.
\eqno(2.7)
$$
The 1-2 entries, $\delta m^2_{12}$, are constrained
\ to be very small compared to $m^2_1$ and $m^2_2$ from FCNC considerations.
Suppose we
put just one of the $\delta m^2_{12}$,
 say ${\delta m^2_{12}}_L$, equal to zero.
 Then, we have a $U(1)$ symmetry acting on the
left-handed lepton superfield of the first generation,
 which will prevent the generation of any Yukawa coupling for the first
generation. Hence, the radiatively generated first
 generation Yukawa coupling will be suppressed relative to the second generation
one by roughly
$$
{\lambda_1\over \lambda_2} \sim {{\delta m^2_{12}}_L \over m^2_L}
 {{\delta m^2_{12}}_R \over m^2_R},
\eqno(2.8)
$$
where the $m^2_{L,R}$ are typical scalar masses
 for the first two generations.

 Let us make
 a more careful estimate for the size of this
suppression. For simplicity, we work
 in the mass insertion approximation where $m^2_1
 , m^2_2, m^2_3$ are taken to be degenerate and
equal to $m^2$.  We find 
the radiatively generated Yukawa matrix for the upper 2 $\times$ 2 block is
$$
\lambda_{2 \times 2} = \left(\begin{array}{ll}
  {{\delta m^2_{12}}_L \over m^2}{{\delta m^2_{12}}_R \over m^2} f(7) x
& {{\delta m^2_{12}}_L \over m^2} f(6) x\\
 {{\delta m^2_{12}}_R \over m^2} f(6) x &
f(5) x\end{array}\right),
\eqno(2.9)
$$
where
$$
f(n) =m^{2n - 4} \int {d^4 k \over {(2 \pi)}^4} {1\over {{(k^2 - m^2)}^{n - 1} 
{(k^2 - M^2)}}} ,\; 
x=\mbox{const}\times 
M {{\Delta m^2_{23}}_L \over m^2}{{\Delta m^2_{23}}_R \over m^2},
$$
and $M$ is the gaugino mass. 
Since $f(n)$ is only logarithmically sensitive to the
ratio ${M^2\over m^2}$, we put $M^2 = m^2$. Then, $f(n) = {1\over{(n-2)(n-1)}}$
and we have
$$
\lambda_{2 \times 2} = \left(\begin{array}{ll}
{1\over {30}}{{\delta m^2_{12}}_L \over m^2}{{\delta m^2_{12}}_R \over m^2} x
& {1\over {20}} {{\delta m^2_{12}}_L\over m^2} x\\
{1\over {20}}{{\delta m^2_{12}}_R \over m^2} x 
& {1 \over {12}} x\end{array}\right).
\eqno(2.10)
$$
Diagonalizing the above matrix, we find  the ratio of the first to second 
generation eigenvalues to be
$$
{\lambda_1 \over \lambda_2} \sim {1 \over {25}} {{\delta m^2_{12}}_L
\over m^2} {{\delta m^2_{12}}_R \over m^2}.
\eqno(2.11)
$$
We see that it is impossible to generate  large enough first generation Yukawa
couplings consistent with FCNC constraints (unless the 
scalars are taken to be unacceptably heavy), which require (for 300 GeV sleptons
and 500 GeV squarks)
$$
\eqalignno{
{1\over{25}} {{\delta m^2_{12}}_\ell \over m^2} {{\delta m^2_{12}}_e \over m^2}
< 2 \times 10^{-4} & (\mu \rightarrow e \gamma)\cr
{1\over {25}} {{\delta m^2_{12}}_q \over m^2} {{\delta m^2_{12}}_d \over m^2}
< 1 \times 10^{-6} & (K_1 - K_2 \; \mbox{mixing})\cr
{1\over {25}} {{\delta m^2_{12}}_q \over m^2} {{\delta m^2_{12}}_d \over m^2}
< 6 \times 10^{-5} & (D_1 - D_2 \;\mbox{mixing}).
&(2.12)\cr}
$$

We are left with the case $K$=1, where Yukawa couplings for 
two generations occur at tree level, while the remaining 
Yukawa couplings, which necessarily correspond to the lightest
generation, are radiatively generated.
In the next section, we study the phenomenological constraints 
on this scenario in detail.

\section{Phenomenological constraints}
\hspace{\parindent} In this section, 
we discuss the phenomenology 
 of obtaining the first
generation Yukawa coupling radiatively. Recall that we are relying on the scalar
mass matrices to break the chiral symmetries associated with the Yukawa
matrices; in particular, then, the scalar mass matrices cannot be diagonalized
in the same basis as the Yukawa matrices. Thus, if we work in the mass
eigenstate basis for all fields, we will have non-trivial mixing matrices at the
gaugino vertices. Let us set the relevant notation here, following \cite{ACH}.
The superpotential contains 
$$
W \supset Q^T \mbox{\boldmath$\lambda_U$} U^c h_u + Q^T 
\mbox{\boldmath$\lambda_D$} D^c h_d + L^T \mbox{\boldmath$\lambda_E$} E^c h_d
\eqno(3.1)
$$
where {\boldmath$\lambda_U$}, {\boldmath$\lambda_D$}, {\boldmath$\lambda_E$} 
are the Yukawa matrices, and are diagonalized by
$$
\begin{array}{c}
\mbox{\boldmath$\lambda_U$} = V_{U_L}^* \overline{\mbox{\boldmath$\lambda_U$}}
 V_{U_R}^{\dagger} \\
\mbox{\boldmath$\lambda_D$} = V_{D_L}^* \overline{\mbox{\boldmath$\lambda_D$}}
 V_{D_R}^{\dagger} \\
\mbox{\boldmath$\lambda_E$} = V_{E_L}^* \overline{\mbox{\boldmath$\lambda_E$}}
 V_{E_R}^{\dagger} . \end{array}
\eqno(3.2)
$$

The soft supersymmetry breaking masses matrices are contained in 
$$
\eqalignno{
\widetilde{Q}^\dagger {\bf{m}}^{2*}_Q \widetilde{Q}
&+ \widetilde{U}^{c\dagger}
{\bf{m}}^2_U
\widetilde{U}^c + \widetilde{D}^{c\dagger} {\bf{m}}^2_D
\widetilde{D}^c + \widetilde{L}^\dagger
{\bf{m}}^{2*}_L \widetilde{L} + \widetilde{E}^{c\dagger}
{\bf{m}}^{2}_E \widetilde{E}^c\cr}
$$
and are diagonalized by
$$
\eqalignno{
{\bf{m}}^{2*}_Q &= U_Q \overline{\bf{m}}^{2*}_Q U^\dagger_Q, \;\;
{\bf{m}}^2_U     = U_U \overline{\bf{m}}^2_U U^\dagger_U, \;\;
{\bf{m}}^2_D     = U_D\overline{\bf{m}}^2_D U^\dagger_D, \cr
{\bf{m}}^{2*}_L     &= U_L\overline{\bf{m}}^{2*}_L U^\dagger_L, \;\;
{\bf{m}}^{2}_E = U_E\overline{\bf{m}}^{2}_E U^\dagger_E ,
&(3.4) \cr}
$$
In the mass eigenstate basis, the rotation matrices $V, U$ appear in the
gaugino couplings,
$$
\eqalignno{
{\cal{L}}_g &= \sqrt{2} g' \sum^4_{\pi = 1}
 \bigg[ -\half \overline{e}_L
W^\dagger_{E_L}\widetilde{e}_L N_n(H_{n\widetilde{B}}
+ \cot \theta_W H_{n\widetilde{w}_{3}}) +
\overline{e}^c_L W^\dagger_{E_R} \widetilde{e}_R
N_n H_{n\widetilde{B}}\cr
&+ \half \cot \theta_W\overline{\nu}_L\widetilde{\nu}_L N_n
H_{n\widetilde{w}_{3}}\cr
&+\overline{u}_L W^\dagger_{U_L} \widetilde{u}_L N_n({1\over 6}
H_{n\widetilde{B}}+\half\cot\theta_W H_{n\widetilde{w}_3})
+\overline{d}_LW^\dagger_{D_L} \wt{d}_L N_n({1\over 6}
H_{n\widetilde{B}}-\half\cot\theta_WH_{n\widetilde{w}_3})\cr
&- {2\over 3} \overline{u}_L^c W^\dagger_{U_R}
 \widetilde{u}_R N_n H_{n\widetilde{B}} +
{1\over 3} \bar{d}^c_L W^\dagger_{D_R}\widetilde{d}_R
N_n H_{n\widetilde{B}} + h.c.\bigg]\cr
&+ g\sum^2_{c=1} [ \bar{e}_L W^\dagger_{E_L} \widetilde{\nu}_L
(\chi_c K_{c\widetilde{w}})
+\bar{\nu}_L \widetilde{e}_L (\chi^\dagger_c K^*_{c\widetilde{w}})\cr
&+ \bar{d}_L W^\dagger_{D_L} \widetilde{u}_{L} (\chi_c K_{c\widetilde{w}})
 + \bar{u}_L
W^\dagger_{U_L}\widetilde{d}_L
(\chi^\dagger_c K^* _{c\widetilde{w}}) +h.c.]\cr
&+\sqrt{2} g_3 [\bar{u}_L W^\dagger_{U_L}\widetilde{u}_L\widetilde{g}
 + \bar{d}_L
W^\dagger_{D_L} \widetilde{d}_L \widetilde{g}
+ \bar{u}^c_L W^\dagger_{U_R}
\widetilde{u}_R\widetilde{g} +
\overline{d}^c_LW^\dagger_{D_R}\wt{d}_R\wt{g} + h.c.],& (3.5)\cr}
$$
here\footnote{Neutrino masses are not discussed here.}
the neutralino and chargino mass
eigenstates are related to the gauge eigenstates by e.g. $\wt{B} =
\sum^4_{n=1} H_{n\wt{B}}N_n,\, \wt{w}_3
= \sum^4_{n=1} H_{n\wt{w}_3} N_n,\,
\wt{w}^+ = \sum^2_{c=1} K_{c\wt{w}}\chi_c$, and
$$
\eqalignno{
W_{E_L} &= U_L^\dagger V_{E_L}, \;
W_{E_R} = U^\dagger_E V_{E_R}, \; W_{U_L}
= U^\dagger_Q V_{U_L},  \; W_{D_L} = U^\dagger_Q V_{D_L},\cr
 W_{U_R} &= U^\dagger_U
 V_{U_R},\; W_{D_R} = U^\dagger_D
V_{D_R}. &(3.6)\cr}
$$

Having defined our notation, we now consider the dominant radiative
contributions to the lepton, up and down mass matrices given in Fig.\ 1. 
In the
following, we assume that the first two generation scalars are
degenerate, since we know from the previous  section that the contribution to
the mass matrix from the non-degeneracy between the first two generations is
negligible. Evaluating the diagrams, we find (keeping only the contribution from
the third generation tree-level mass) \cite{HKR} :
$$
\eqalignno{
\Delta {\bf m_e}_{\, \alpha \beta} = & 
\sum^4_{n =1} {H_{n\tilde{B}}\over M_n}(H_{n\tilde{B}} + \cot
\theta_W H_{n\tilde{w}_3}) \cr 
& \times {{\alpha m_{\tau}}\over {4 \pi \cos^2 \theta_W}} 
(A + \mu \tan\beta) \times 
\{ W_{E_L 3 \alpha} W^*_{E_L 33} \, W_{E_R 3 \beta} W^*_{E_R 33} \cr
 & [ h(x_{3L_n},x_{3R_n}) 
- h(x_{3L_n},x_{1R_n}) -   h(x_{1L_n},x_{3R_n}) + 
h(x_{1L_n},x_{1R_n}) ]  \cr
 & + W_{E_L 3 \alpha} W^*_{E_L 33} \, \delta_{3 \beta} [
h(x_{3L_n},x_{1R_n})  -  
h(x_{1L_n},x_{1R_n}) ] \cr
 & + \delta _{\alpha 3} \, W_{E_R 3 \beta} W^*_{E_R 33}[ 
h(x_{1L_n},x_{3R_n}) -  
h(x_{1L_n},x_{1R_n})]  \cr
 & + \delta_{\alpha 3}  \delta_{3 \beta} 
 h(x_{1L_n},x_{1R_n}) \} ,
& (3.7) }
$$
$$
\eqalignno{
\Delta {\bf m_u}_{\, \alpha \beta} = 
& {8\over 3}{{\alpha_s m_t}\over {4 \pi}} 
({{A + \mu \cot\beta}\over {M_{\tilde {g}}}}) \times
 \{ W_{U_L 3 \alpha} W^*_{U_L 33} \,W_{U_R 3 \beta} W^*_{U_R 33}\cr
 & [ h(x_{3L},x_{3R}) 
- h(x_{3L},x_{1R}) -   h(x_{1L},x_{3R}) + 
h(x_{1L},x_{1R}) ]  \cr
 & + W_{E_L 3 \alpha} W^*_{E_L 33}  \delta_{3 \beta} [
h(x_{3L},x_{1R})  -  
h(x_{1L},x_{1R}) ]
\cr
 &  + \delta _{\alpha 3} \, W_{E_R 3 \beta} W^*_{E_R 33}[ 
h(x_{1L},x_{3R}) - 
h(x_{1L},x_{1R})]
 \cr
 & + \delta_{\alpha 3} \, \delta_{3 \beta} \,
 h(x_{1L},x_{1R}) \}, 
& (3.8) }
$$
where $x_{3L(R)_n} = {{m^2_{\tilde{\tau}_L(R)}}\over M^2_n}$,
$x_{1L(R)_n} = {{m^2_{\tilde{e}_L(R)}}\over M^2_n}$ in the 
lepton sector, $x_{3L(R)} = {{m^2_{\tilde{t}_L(R)}}\over M^2_{\tilde{g}}}$,
 $x_{1L(R)} = {{m^2_{\tilde{u}_L(R)}}\over M^2_{\tilde{g}}}$   
and $\Delta {\bf{m_d}}_{\, \alpha \beta}$ is the same as
$\Delta {\bf {m_u}}_{\, \alpha \beta}$ with the replacements 
$\cot\beta \rightarrow \tan\beta, m_t 
\rightarrow m_b$ and $\wt{t},\wt{u} \rightarrow \wt{b},\wt{d}$,
and where
$$
\eqalignno{
h(x,y) &= {{f(x) - f(y)}\over {x - y}}, \cr
f(x) &= {{x \ln x}\over{1-x}}.
&(3.9)}
$$

Let us begin our phenomenological discussion with the lepton sector. The above
expression for the radiative contribution to the lepton mass matrix is rather
unwieldy; while we can use it for numerical work, in order to get an approximate
feeling for the size of the radiative electron mass, we simply look at the 11
entry of the radiative correction matrix $ m_e \approx
\Delta {{\bf m}_e}_{11}$. For simplicity, we assume that one of the
neutralinos is pure bino, that the scalar tau's are degenerate with mass $m$ and
much lighter than the selectrons. Then we find as in \cite{ACH1}
$$
m_e = {{\alpha m_{\tau}}\over {4 \pi \cos^2 
\theta_W}} {{(A + \mu \tan\beta)}\over{M_1}} \times 
W_{E_L 31} W_{E_R 31} h(x_3,x_3),
\eqno(3.10)
$$
where $M_1$ is the bino mass, 
$h(1,1) = 1/2$, and
we have assumed $W_{E 33} \simeq 1$.
 As explained in \cite{ACH1}, we must work in the large
$\tan\beta$ regime, and so we can neglect 
the $A$ term contribution above.
If we set $\tan\beta = 60$ and $\mu = M_1 = m$, equation (3.10) 
reproduces the
electron mass if the product $\wer_{31} \wel_{31} \simeq 0.01$. This 
is roughly speaking a lower bound for this product. In this calculation we 
have taken the selectron to be much heavier than the stau so that the super-GIM
cancellation in the loop can be ignored. In fact, however, for selectrons
moderately heavier than the staus, there will be a super-GIM cancellation and   
$\wer_{ 31} \wel_{ 31}$ will be correspondingly larger. In Fig. 2,
we give a plot for the relevant super-GIM suppression factor. 
Assuming  left and right handed scalars
degenerate, scalars of the first two generations degenerate,
 and the third generation scalar degenerate with the  gaugino, 
we plot the super-GIM factor against the ratio of first two generation 
to the third generation scalar
masses. This implies that 
each of $\wer_{31}$,$ \wel_{ 31}$ should be at least 0.1. In the following we
will explore the
consequences of having such large mixing angles. 

--$\mu\to e\gamma$: One immediate observation is
that, if in the diagram of  Fig.\ 1(a) we replace 
one of the external electrons with
a muon and attach  a photon to the graph, we get a potentially dangerous
contribution to the rare process $ \mu \rightarrow e \gamma$. How dangerous is
this effect? In appendix B,
we present the FCNC constraints on the elements of the mixing matrices
$W$.  Requiring the  $\mu \rightarrow e \gamma$ rate to be smaller than current
experimental bound constrains $\welr_{32} \werl_{31}$ to be smaller than $\sim
10^{-4}$. Since we know that we need $\welr_{31} \sim 0.1$ in order to
generate the electron mass radiatively, we must have that $\welr_{32} \lsim
10^{-3}$ in order to avoid a dangerous $\mu \rightarrow e \gamma$ rate. It may
seem
strange that $W_{E_{L(R)} 31}$ and $W_{E_{L(R)} 32}$ 
have such disparate sizes;
any theory of lepton flavor with radiatively generated electron mass must
naturally explain why $W_{E_{L(R)}32}$ is so much smaller than $W_{E_{L(R)}
31}$.  Speaking more loosely, 
if the electron mass is radiative, muon number must
be very nearly conserved. 

--$\tau\to e\gamma$: What about the decay $\tau \rightarrow e \gamma$? 
Since it is a 3-1 transition,
it is directly related to $\welr_{31}$. Under the same set of assumptions
that went into the simplified equation (3.10), 
the amplitude for $\tau_{L(R)}$ decay
is
$$
F_{L(R)} =  {{\alpha m_{\tau}}\over {4 \pi \cos^2 
\theta_W}} {{(A + \mu \tan\beta)}\over{M^3_1}} \times 
W_{E_{L(R)} 31}  g (x_3,x_3),
\eqno(3.11)
$$
where 
$$
\eqalignno{
g(x,y) &= {{f^{\prime}(x) - f^{\prime}(y)}\over {x - y}}, \cr
f^{\prime}(x) &= {{x^2 - 2 x \ln x - 1}\over {2(x - 1)^3}},
&(3.12)}
$$
and  $g(1,1) = {1\over 12}$. The branching ratio for $\tau
\rightarrow e \gamma$ is proportional to $|\wel_{31}|^2 + |\wer_{31}|^2 \geq 
2 |\wel_{31} \wer_{31}|$, which is the 
product constrained by the requirement of 
obtaining radiative electron mass.  Putting $\mu= M_1 = m =$ 300 GeV gives 
$B(\tau \rightarrow e \gamma) \approx 10^{-6}$, a factor of 100 beneath the
current bound. We  make a more careful analysis as follows. Assuming
that the left and right scalars, as well as the scalars of the first two
generations  are degenerate, both the radiatively generated $m_e$ and the $\tau
\rightarrow e \gamma$ rate depend on the following parameters (other than the
mixing angles) in the large
$\tan\beta$ regime: $(\mu,M_1,M_2,m^2_{\tilde{\tau}},m^2_{\tilde{e}},
\tan \beta)$
Putting $\tan \beta =60$ and assuming the grand unification 
relation $M_2 \sim 2 M_1$,
the dependence is reduced to only $(\mu, M_1,m^2_{\tilde{\tau}},
m^2_{\tilde{e}})$. Specifying these parameters determines what the product   
$|W_{E_L 31} W_{E_R 31}|$ should be to obtain the correct electron mass, and
this in turn provides us with a lower bound on $B(\tau \rightarrow e \gamma)$.
In Fig.\ 3, we give a representative contour plot 
for this lower bound on $B(\tau \rightarrow e \gamma)$. Over a
significant portion of the parameter space, the rate is only  10-100 times
smaller than  the current bound 
$B(\tau \rightarrow e \gamma) \lsim 1.2\times 10^{-4}$
\cite{tau}.

--$d_e$: If there are $CP$ violating phases in the theory, we have further
considerations. First, we note that if there is no mixing with 
the second generation (as seems to be required for avoiding
dangerous $\mu \rightarrow e \gamma$), then we can choose a basis
where the mixing matrices $\welr$ are real: the only potentially complex
coupling is 
$(\tilde{e}_L^{*} m^2_{13 L} \tilde{\tau}_L + h.c. + L \rightarrow R$).
Since the tree level electron Yukawa coupling is zero, we can 
independently rephase the superfields $e_{L(R)}$ to make 
$m^2_{13 L(R)}$ real. Thus, the only sources of $CP$ violation
are the phases in the $A$ and $\mu$ parameters.  Ordinarily,
(when no fermion masss are generated radiatively), the phases of
$A$ and $\mu$ are constrained to be small, since arbitrary phases lead to
large electric dipole moments via diagrams proportional to the tree
level first generation Yukawa couplings. Does the situation change
when we generate the lightest generation Yukawa coupling 
radiatively? To answer this question, let us 
look at the lepton mass matrix and dipole moment matrix in the
2 dimensional space of the first and third generation (the second
generation has no mixing and is thus irrelevant). For
simplicity, we again consider taking  the first 
two generation scalars much heavier than
those of the third generation so that they are
decoupled, and we set $\mu = M_1 = m$. Then, we have
$$
\eqalignno{
{{\bf{m}_e}\over m_{\tau}}
& \simeq \left(\begin{array}{ll}
.02 \,\wel_{31} \wer_{31} e^{i \theta} & .02 \, \wel_{31} e^{i \theta} \\
.02 \, \wer_{31} e^{i \theta} & 1\end{array}\right), \cr
\cr
{{\bf{d}_e}\over e}  & \simeq 1.5 \times 10^{-21} cm \times 
 {({{300 \mbox{GeV}}\over {M_1}})}^2 \cr
 & \times 
\left(\begin{array}{ll} \wel_{31} \wer_{31} & \wel_{31} \\
\wer_{31} & 1 \end{array}\right)
e^{i \theta}, &(3.13)} 
$$
where $\theta$ is the phase of $A + \mu \tan \beta$. We can 
approximately diagonalize the lepton mass matrix as follows
$$
\eqalignno{
{{\bf{m}_e}\over m_{\tau}} & \simeq V^{*}_{E_L}
\left(\begin{array}{ll} .02 \, \wel_{31} \wer_{31}
& 0\\
0 & 1\end{array}\right)
V^{\dagger}_{E_R}, \cr
\cr
V_{E_{L(R)}} & \simeq
\left(\begin{array}{ll} e^{-i \theta/2} & .02 \, \welr_{31} e^{-i \theta}\\
-.02 \, \welr_{31} e^{i \theta/2} & 1\end{array}\right).
&(3.14)} 
$$
In the basis where the lepton mass matrix is diagonal with 
real eigenvalues, the electric
dipole moment matrix is ${\bf{d}}^{\prime}_e = V^T_{E_L} 
{\bf{d}}_e V_{E_R}$, and the electric dipole moment of the electron
is $d_e = Im({{\bf{d}}^{\prime}_e}_{11})$. We find with $M_1 =$
300 GeV and  $ \wel_{31} \wer_{31} \sim .01$ (as required
to generate the electron mass),
$$
{d_e\over e}
 =  6\times 10^{-24} \mbox{cm}\times \sin \theta.
\eqno(3.15)
$$ 
Thus, $\sin \theta$ must be smaller than $\sim 7 \times 
10^{-4}$ for ${d_e\over e}$ not to exceed the experimental 
limit of $4 \times 10^{-27}$ cm.   So, we have not made
any progress on the supersymmetric
$CP$ problem. However, as we have already mentioned, if
we assume that $\sin \theta$ is sufficiently suppressed, there
are no other $CP$ violating contributions 
when muon number is conserved. 

What if the electron mass is not all radiative 
in origin and has 
some small tree level contribution? If there
is an $O(1)$ phase mismatch between the
tree and radiative parts of the electron mass,
there will be a  phase in the electron 
electric dipole moment  of order 
${{m^{tree}_e}\over {m_e}}$ even if $A$ and
$\mu$ are taken to be real. This would again
give too large a dipole moment unless 
${{m^{tree}_e}\over {m_e}} \lsim 10^{-3}$. 
(Of course, in deriving this result,we assume that
most of the electron mass is radiative, 
otherwise there is no reason for  
the $\welr_{31}$ to be big enough to cause trouble
with the dipole moment). We conclude that if there are large 
$CP$ violating phase differences
in the theory, the electron mass must either be nearly all
radiative or nearly all tree level.  

In the quark sector, in addition to the first generation quark masses,
we are also interested in the possibility of generating CKM mixing
angles by  finite radiative corrections. Table 1 shows the relevant
ratios of quark masses and mixing angles.
\begin{table}[htb]
\begin{tabular}{||c|c||c|c||c|c||}  \hline
${m_t\over m_t}$ & 1 & ${m_b\over m_t}$ & $1.6\times 10^{-2}$ &
${m_b\over m_b}$ & 1 \\ \hline
${m_c\over m_t}$ & $3.6\times 10^{-3}$ & ${m_s\over m_t}$ &
$4.5\times 10^{-4}$ & ${m_s\over m_b}$ & $2.7\times 10^{-2}$ \\ \hline
${m_u\over m_t}$ & $1\times 10^{-5}$ & ${m_d\over m_t}$ & $2\times 10^{-5}$ &
${m_d\over m_b}$ & $1.3\times 10^{-3}$ \\ \hline
${\sin\theta_c \,m_c\over m_t}$ & $8\times 10^{-4}$ & 
${\sin\theta_c \, m_s\over m_t}$ & $1\times 10^{-4}$ & 
$ {\sin\theta_c \, m_s\over m_b}$ & $6\times 10^{-3}$ \\ \hline
${V_{cb}\, m_t\over m_t}$ & $4\times 10^{-2}$ & ${V_{cb}\, m_b\over m_t}$ &
$6.4\times 10^{-4}$ & ${V_{cb}\, m_b\over m_b}$ & $4\times 10^{-2}$ \\ \hline
${V_{ub}\, m_t\over m_t}$ & $4\times 10^{-3}$ & ${V_{ub}\, m_b\over m_t}$ &
$6.4\times 10^{-5}$ & ${V_{ub}\, m_b\over m_b}$ & $4\times 10^{-3}$ \\ \hline
${V_{td}\, m_t\over m_t}$ & $1\times 10^{-2}$ & ${V_{td}\, m_b\over m_t}$ &
$1.6\times 10^{-4}$ & ${V_{td}\, m_b\over m_b}$ & $1\times 10^{-2}$ \\ \hline
\end{tabular}

\vskip .3in

{\bf Table 1}: The relevant ratios of quark masses and mixing angles with all 
quantities taken at the scale of top quark mass. The values of
quark masses, mixing angles, and the RG mass enhancement factors $\eta_i$
are taken as follows: $m_t(m_t)=168\,\gev$, $m_b(m_b)=4.15\,\gev$,
$m_c(m_c)=1.27\,\gev$, $m_s(1\,\gev)=180\,\mev$, $m_d(1\,\gev)=8\,\mev$,
$m_u(1\,\gev)=4\,\mev$, $\eta_b=1.5$, $\eta_c=2.1$, $\eta_{u,d,s}=2.4$,
$\sin\theta_c=0.22$, $V_{cb}=4\times 10^{-2}$, $V_{ub}=4\times 10^{-3}$,
$V_{td}=1\times 10^{-2}$.
\end{table}

The constraints on SUSY FCNC have been studied in \cite{HKT,GMS}, 
and the results are given in terms of $\delta_{ij}={{\delta \tilde{m}^2_{ij}}
\over M^2_{\tilde{q}}}$, where $\delta \tilde{m}^2_{ij}$ is the off diagonal
squark mass in the super-KM basis and $M_{\tilde{q}}$ is the ``universal
squark mass''. However, in order to generate the light  generation quark 
masses entirely by radiative corrections, the splitting between  scalar masses 
of the first two and the third generations must be quite large 
so that the super-GIM cancellation is not effective. As we
can see from Fig.\ 2, this typically 
requires ${\tilde{m}_1 \over \tilde{m}_3} \gsim 3$.
Then it is not clear which scalar mass should be used for $M_{\tilde{q}}$.
In appendix B, we translate thes results obtained in \cite{HKT,GMS} into
constraints directly on the mixing matrix elements, which are more
suitable for our dicussions. 

When \tb\ is large, some of the one-loop diagrams for the down type quark
Yukawa couplings are enhanced by \tb\ (Figs.\ 1(c), 4(a)(b)).
They can give significant corrections to the down type quark masses
and CKM matrix elements\cite{BRP}. Here we are interested in the possibility
that some of the light generation quark masses and mixing angles are 
entirely generated by these loop corrections. Because of the large \tb\
enhancement, it is easier to generate CKM mixing angles in the down
sector than in the up sector. In fact, we can see from Table 1
that it is impossible to generate \vcb\ in the up sector, while generating
\vub\ and $\theta_c$ requires $\wul_{31}$ to be greater than about
0.4 and 0.2 respectively. $\wul$ is linked to $\wdl$ by the CKM matrix:
$V_{CKM}=\wul^\dagger \wdl$. To get the correct \vub, $\wul_{31}$ has to
be canceled by the mixing angles of the same size in $\wdl$, which will
violate the FCNC constraints listed in Table B1. Therefore, we will
only consider generating CKM mixing angles in the down sector. 

The flavor diagonal gluino diagram  could give large 
corrections to the down quark masses if the corresponding Yukawa
couplings already exist at tree level. It does not generate fermion
masses if they are absent at tree level, but  gives 
large uncertainties in the tree level bottom Yukawa
coupling $\lambda_b^0$, which appears in these gluino diagrams. 
The flavor-changing gluino diagram (through $m_b^0 \mu \tb$) can
give sizable down quark mass matrix elements involving light generations
and therefore generate $m_d$ and CKM mixing angles. The first
chargino diagram (Fig.\ 4(a)) 
only gives significant contributions when one of the
external leg is $b_R$, i.\ e., it contributes to ${\lambda_D}_{13},\;
{\lambda_D}_{23},\;{\lambda_D}_{33}$. With some unification assumptions
at high scales, one usually finds the chargino contribution to the 
bottom quark mass is smaller than and opposite to the gluino contribution
\cite{COPW,HRS}. Here we do not make assumptions about physics at high scales
so both contributions lead to uncertainties in the tree level $\lambda_b^0$.
The contributions to ${\lambda_D}_{13}$ and ${\lambda_D}_{23}$ are 
proportional to \vtd\ and \vts\ respectively, so they can only give corrections
to the already existing mixing angles but not generate them entirely.
The second chargino diagram (Fig.\ 4(b)) is supressed by the weak coupling
constant compared with other diagrams and will be ignored. In the following
we will concentrate on the possibilities that the light fermion masses
and mixing angles are generated by the flavor-changing gluino diagram.

--$m_u$: The possibility that $m_u$ comes from radiative corrections by
mixing with the third generation has been pointed out in \cite{ACH}.
We can see from Fig.\ 2 and ${m_u \over m_t}$ in Table 1 that if 
$\wul_{31} \wur_{31} \sim 10^{-3}$, $m_u$ can be generated entirely from
radiative corrections. There is no direct constraint on the 1-3 mixing.
The induced splitting between the first two generation left-handed
squark masses could contribute to $K-\bar{K}$ mixing. However, this
constraint is easily satisfied, so it is possible that $m_u$ is 
entirely radiative. 
 
--$m_d$: From Fig.\ 2 and Table 1. we can see that to generate $m_d$
requires $\wdl_{31} \wdr_{31} \sim 2\times 10^{-3}$. Compared with
the constraints derived from $B-\bar{B}$ mixing in Table B1(a), 
this requires
the sfermion masses to be in the TeV range, which is somewhat
uncomfortably large. In addition, if $m_d$ does get its mass from
radiative corrections, we also generate the 1-3 entry for the down
Yukawa matrix. Their ratio is:
$$
{{\Delta {\lambda_D}_{11}}\over {\Delta {\lambda_D}_{13}}} =
{{\wdl_{31}\wdr_{31}\wt{\wt{H}}}
\over {\wdl_{31}\wdr_{33}\wt{H}}}
< {\wdr_{31}\over \wdr_{33}} \lsim 0.1 , \eqno{(3.16)}
$$
for $m_{\tilde{b}} \sim 1$TeV, assuming $\wdr_{33} \simeq 1$,
where $\wt{\wt{H}}=h(x_{3L},x_{3R})-h(x_{3L},x_{1R})-h(x_{1L},x_{3R})
+h(x_{1L},x_{1R}),\; \wt{H}=h(x_{3L},x_{3R})-h(x_{1L},x_{3R})$,
and $h$, $x_{1(3)L(R)}$ are defined in (3.8), (3.9).
On the other hand, ${m_d \over {V_{ub} m_b}} \simeq 0.3$.
We see that the generated $\Delta {\lambda_D}_{13}$ gives a too big
contribution to \vub\, which has to be canceled by a tree level 
${\lambda_D}_{13}$.

We now discuss the possibilities for radiative generation
of CKM elements. We take the independent parameters of the
CKM matrix to be $V_{us}, V_{ub}, V_{cb}$ and the $CP$ 
violating phase.

--$\theta_c$: To generate $\theta_c$ we need $\wdl_{31}\wdr_{32}
\sim 10^{-2}$, assuming $\wdlr_{33}\simeq 1$. 
From $B-\bar{B}$ mixing and $b\to s\gamma$ decay,
or $K-\bar{K}$ mixing alone, the sfermion masses are also required
to be $\gsim 1$TeV in order to satisfy these constraints.
Furthermore the phase of $\wdl_{31}\wdr_{32}$ has to be small ($< 10^{-1}$)
from the $\epsilon$ parameter of CP violation. Similar to the case of
$m_d$, generating $\theta_c$ radiatively may also give 
a too big contribution
to \vub.
If we try to generate $m_d$, $\theta_c$, and \vub\ all by radiative
corrections, ignoring the difference between $\wt{H}$ and $\wt{\wt{H}}$,
we obtain the following ratio for the mixing matrix elements from
Table 1:
$$
\wdr_{33}\: : \: \wdr_{32}\: :\: \wdr_{31}\; \simeq \;
V_{ub}m_b\; : \sin\theta_c m_s\; :\; m_d\; \simeq \;
4\; :\; 6\; :\; 1.3 . \eqno{(3.17)}
$$
By unitarity we obtain
$$
\wdr_{33}\simeq 0.55,\; \wdr_{32}\simeq 0.82,\; \wdr_{31}\simeq 0.18.
\eqno{(3.18)}
$$
(Taking into account that $\wt{H} > \wt{\wt{H}}$ gives larger $\wdr_{32}$,
$\wdr_{31}$.) From Table B1, we can see that $m_{\tilde{b}}$ has to be pushed
above 2 TeV (even higher for the first two generations) to satisfy the 
constraints from both $\Delta M_K$ and $b\to s\gamma$. If there are 
{\cal O}(1) phases in these $W$'s, the $\epsilon$ constraints raise the 
lower limit of the squark masses to $\sim$ 20 TeV, which is unacceptably
large. Furthermore, it is unnatural for models to have such a large
$\wdr_{32}$ mixing. 
Therefore, it is unlikely that all CKM matrix elements can be generated 
by radiative
corrections. 

--$V_{ub}$: To generate \vub\ we need $\wdl_{31} \sim 5\times 10^{-3}$, which
easily satisfies the $B-\bar{B}$ mixing constraints. Hence \vub\ can
be generated radiatively, but as we learned from above, \vub\ and $\theta_c$
cannot both come from radiative corrections, and neither can \vub\ and 
$m_d$.

--$V_{cb}$: Attaching a photon to the diagram which generates 
$\Delta {m_D}_{23}$ gives a diagram contributing to the 
decay $b \to s \gamma$ . Hence one can write down the
 following simple relation between
gluino diagram contributions to \vcb\ and to the Wilson coefficient
$c_7(M_W)$ \cite{c7} for $b\to s \gamma$,
$$
\Delta c_7(M_W) = q_D \,{{4\pi}\over \alpha} \sin^2 \theta_W
{M_W^2\over m_{\tilde{g}}^2}\, {\wt{G}\over \wt{H}} \,{\Delta {m_D}_{23} \over
V_{cb} m_b}, \eqno{(3.19)} 
$$
$$
\Rightarrow {\eta^{16/23} \Delta c_7(M_W) \over 
c_7(m_b)_{\mbox{\scriptsize{SM}}}}
\simeq ({8m_W\over m_{\tilde{g}}})^2 \,({5\wt{G}\over \wt{H}}) \,
({\Delta {m_D}_{23} \over V_{cb} m_b}). \eqno{(3.20)}
$$
where $\wt{G}= g(x_{3L},x_{3R})-g(x_{1L},x_{3R})$, and $g$ is defined in
(3.12).
The gluino diagram contribution to $b \to s\gamma$ interferes constructively
with
the Standard Model contribution if \vcb\ is generated by the similar
gluino diagram. Therefore, generating \vcb\ radiatively requires
heavy gluino and squark masses ($\gsim 1$ TeV) or  cancellation 
between the chargino diagram 
contributions to $b \to s \gamma$ and other contributions.

--CP-violating phases: From the above discussion we found that it is
very difficult to generate all CKM mixing matrix elements by radiative
corrections. This means that a non-trivial CKM matrix should 
occur
at  tree level. There is one physical CP-violating phase in $V_{CKM}$,
and several more in the 
quark-squark-gaugino mixing matrices. The number of CP-violating phases
in the quark sector (not including the possible phases of the parameters
$A$ and $\mu$) is counted as in the following. There are four unitary
mixing matrices $\wul ,\, \wur ,\, \wdl ,\, \wdr,$ ($V_{CKM}=\wul^{\dagger}
\wdl$ is not independent,) connecting 7 species of quark and squark
fields $u_L,\, d_L,\, u_R,\, d_R,\, \wt{Q},\, \wt{U},\, \wt{D}$.
Among the phases of these fields, 6 are fixed by the 
6 eigenvalues of the Yukawa matrices
$\mbox{\boldmath{$\lambda$}}_U$ and $\mbox{\boldmath{$\lambda$}}_D$ 
(if there are no zero eigenvalues), one
overall phase is irrelevant, so we can remove 14 of the 24 phases in the
$W$'s by phase redefinition of the quark and squark fields. Each massless
quark removes one more phase by allowing independent phase rotations on 
the left and right quark fields. Each pair of degenerate quarks or squarks
of the same species removes one phase as well. Assuming  $m_u$
and $m_d$ massless at the tree level, and degeneracies between
the first two generation squarks, we can remove 5 more phases and there
are still 5 independent phases left. One of them cannot be moved
to the $W_U$'s and it can give significant contributions to the CP
violation effects in the $K$ and $B$ systems.

\section{Guidelines for model building}

\hspace{\parindent} In the introduction 
and in \cite{ACH1} we indicated some general
features effective theories of flavor should have in order to generate radiative
fermion masses. In particular, we pointed out that, in supersymmetric theories,
 an accidental superpotential symmetry is needed to ensure that the first
generation is massless at tree-level, while this symmetry must be broken by $D$
terms in order to obtain radiative masses. For instance, in the effective lepton
models considered in \cite{ACH1}, all holomorphic and flavor symmetric operators
possess an accidental $U(1)_{{\ell}_1} \times U(1)_{e_1}$ which is violated by
the $D$-terms. From the point of view of an effective theory, then, it is
representation content and holomorphy which are responsible for accidental
symmetries for every possible superpotential operator, thereby forbidding some
Yukawa couplings. However, this is by no means a necessary condition for the
existence of tree level massless fermions: 
We do not always generate every operator
consistent with symmetries when we integrate out heavy states. Thus, the
condition that every effective operator in the superpotential possess an
accidental symmetry is clearly too strong; we only need an accidental symmetry
to exist for those operators induced by integrating out heavy states. For this
reason, it seems that a deeper understanding of the accidental symmetries lies
in examining the full theory, including superheavy states. This is our purpose
in this section. We will find simple, sufficient conditions for guaranteeing the
existence of tree level massless states after integrating out heavy states. We
will also describe (in view of later application to the quark sector) the
structure of the tree-level CKM matrix. These conditions will serve as
convenient guides for the explicit models we construct in the next section.

We begin by considering sufficient conditions for the existence of
tree level massless states. Consider the lepton sector for simplicity. In 
Froggatt-Nielsen schemes, we have fields $\ell_\alpha , e_\alpha \; 
(\alpha=1,2, 3)$ 
which would be the 3 low 
energy left and right handed lepton fields in the flavor symmetric limit.
However, there are also superheavy states with which $\ell$ and $e$ mix after
flavor symmetry breaking. In general, we  have  vector-like superheavy
states $(L_i \oplus \bar{L}_i)$ and  $(E_a \oplus \bar{E}_a)$,
($i=4,..., n+3, \; 
a=4,..., m+3$), 
with $L$, $E$ having the same gauge quantum numbers as $\ell, e$ respectively,
and with the barred fields having conjugate gauge quantum numbers.
We also have a set
of gauge singlet fields $\phi$ with VEV's $\VEV{\phi}$ breaking the flavor group
$G_f$. In the superpotential, we have bare mass terms for the $(L,\bar{L})$ and
the $(E,\bar{E})$ fields, as well as trilinear couplings mixing $\phi$'s with
light and superheavy states. We also have a large Yukawa matrix 
${\mbox{\boldmath $\Lambda$}}_{IA} \; (I=1,..., n+3,\, A=1,..., m+3)$,
connecting the down-type Higgs $h_d$
to the $(\ell_\alpha,L_i)$ and $(e_\alpha,E_a)$,
$$
W \supset \left(\begin{array}{ll}\ \ell & L\end{array}\right)
\mbox{\boldmath $\Lambda$} 
\left(\begin{array}{l}\ e \\ E \end{array}\right)
h_d.
\eqno(4.1)
$$
Once the fields $\phi$ develop VEV's, we will have mass terms like, $\ell 
\VEV{\phi} \bar{L}$ mixing light and heavy states. In order to diagonalize the
bare mass matrix and go from the flavor basis to the mass basis (where ``light''
and ``heavy'' are correctly identified), we must make appropriate 
$\VEV{\phi}$ dependent
unitary rotations on the fields:
$$
\left(\begin{array}{l}\ \ell^{\prime} \\ L^{\prime} \end{array}\right) 
= U_L (\VEV{\phi}) \left(\begin{array}{l}\ \ell \\ L\end{array}\right), \;
\bar{L}^{\prime} = U_{\bar{L}}(\VEV{\phi}) \bar{L},
$$
$$ 
\left(\begin{array}{l}\ e^{\prime} \\ E^{\prime} \end{array}\right)
= U_E (\VEV{\phi}) \left(\begin{array}{l}\ e \\ E \end{array}\right),\;
\bar{E} ^{\prime} = U_{\bar{E}}(\VEV{\phi}) \bar{E}.
\eqno(4.2)
$$
In this basis, the mass terms are $\sum_{i=4}^{n+3} M_i \bar{L}_i^{\prime}
 L_i^{\prime}\ + 
\sum_{a=4}^{m+3} M_a \bar{E}_a^{\prime} E_a^{\prime}\ $, and the Yukawa matrix
becomes
$$
\mbox{\boldmath $\Lambda$}_{IA}^{\prime}
 = U_L^*(\VEV{\phi})_{IJ} \mbox{\boldmath $\Lambda$}_{JB} U_E^{\dagger}
(\VEV{\phi})_{BA},
\eqno(4.3)
$$
where summation over $J$ and $B$ is understood.
In order to integrate out the (now correctly identified) heavy states at tree
level, we simply throw out any coupling involving them. The Yukawa matrix
{\boldmath $\lambda$} for
the three low energy generation leptons is then 
$$
\mbox{\boldmath$\lambda$}_{\alpha\beta}
= U_L^*(\VEV{\phi})_{\alpha J} \mbox{\boldmath $\Lambda$}_{JB} 
U_E^{\dagger}(\VEV{\phi})_{B\beta}, \; (\alpha,\beta =1,2, 3).
\eqno(4.4)
$$

We would now like to understand circumstances under which we can guarantee a
certain number of zero eigenvalues for {\boldmath$\lambda$}. For
{\boldmath$\lambda$} to have $k \leq 3$ zero eigenvalues, its rank must be 
$3-k$.
To see when this is possible, we make the simple observation that each row (or
alternatively each column) of {\boldmath$\Lambda$} contributes one rank  to
{\boldmath$\lambda$}. Consider for instance the contribution to
{\boldmath$\lambda$} from the $J_0$'th row of {\boldmath$\Lambda$}. Defining
$$
x_\alpha = U^*_{L_{\alpha J_0}}, \; y_\beta= {\mbox{\boldmath$\Lambda$}}_{J_0 B}
U^\dagger_{E_{B \beta}},
$$
we have
$$
\mbox{\boldmath$\lambda$}_{\alpha \beta}^{\mbox{\scriptsize{from row}}\, J_0} 
=x_\alpha y_\beta,
\eqno(4.5)
$$
which is manifestly rank 1. Define a non-zero
row (column) of {\boldmath$\Lambda$} 
to be a row (column) with at least one non-zero entry.
Then,  it is clear that a {\em sufficient\/} 
condition for   
{\boldmath$\lambda$} to have rank $3 - k$ is that the number of non-zero rows 
(or the number of non-zero columns)  of {\boldmath$\Lambda$}, up to
rotations, equal $3 - k$, i.\ e., {\boldmath$\Lambda$} also has rank 
$3-k$; since
in this case  {\boldmath$\lambda$} is of the form
$$
\mbox{\boldmath$\lambda$}_{\alpha\beta} = \sum_{J=1}^{3-k} x^J_\alpha 
y^J_\beta,
\eqno(4.6)
$$
which is manifestly rank $3 - k$
(the case of interest to us is $k=1$). We will make use of this
criterion in the following section.

We next turn to examining  the tree-level CKM matrix in the quark sector.
In analogy to the lepton sector, we have Yukawa matrices {\boldmath$\Lambda_D$}
and {\boldmath$\Lambda_U$},
$$
W \supset \left(\begin{array}{ll}\ q & Q\end{array}\right)
\mbox{\boldmath $\Lambda_U$} 
\left(\begin{array}{l}\ d \\ D \end{array}\right) h_d
+ \left(\begin{array}{ll}\ q & Q\end{array}\right)
\mbox{\boldmath $\Lambda_D$}
\left(\begin{array}{l}\ u \\ U \end{array}\right) h_u,
\eqno(4.7)
$$
where all new fields are in obvious
analogy with the lepton sector. Let us assume that the  general condition stated
above,
ensuring the existence of a massless eigenvalue for {\boldmath$\lambda_D$} and
{\boldmath$\lambda_U$}, is realized by {\boldmath$\Lambda_D$} and 
{\boldmath$\Lambda_U$}. Then, we can write
$$
{\mbox{\boldmath$\lambda_D$}}_{\alpha\beta} = x_{\alpha}^1 y_\beta + 
x_\alpha^2 z_\beta, \;
{\mbox{\boldmath$\lambda_U$}}_{\alpha\beta} = x_{\alpha}^{\prime 1} 
y_\beta^{\prime} + 
x_\alpha^{\prime 2} z_\beta^{\prime},
\eqno(4.8)
$$
Suppose in particular that {\boldmath$\Lambda_D$} and 
{\boldmath$\Lambda_U$} have nontrivial entries in the same two rows, in which 
case we can choose $x_{\alpha}^i = x_{\alpha}^{\prime i}, \; i=1,2$. 
Then,  the resulting CKM matrix  has non-zero entries only in
the 2-3 sector. The reason is that, since the first generation is massless,
we can always choose a basis where the first generation quark doublet
has no component of superheavy quark doublets with Yukawa couplings, and so 
both {\boldmath$\lambda_D$} and {\boldmath$\lambda_U$} are only non-zero in 
the lower $2\times 2$ block. We can see this more explicitly as follows. 
First note that we can make a rotation on the left handed quarks to
make $x_{\alpha}^1$ point in the 3 direction, and make independent rotations on
the right-handed up and down quarks to make $y_{\beta}$ and $y_\beta^{\prime}$
also point in the 3 direction. In this basis, we have
$$
\mbox{\boldmath $\lambda_D$}_{\alpha\beta}
 = \left(\begin{array}{lll} 0& 0& 0\\ 0 & 0 & 0
\\0 & 0 & 
\eta \end{array}
\right)_{\alpha\beta} + x_\alpha^2 z_\beta, \;
\mbox{\boldmath $\lambda_U$}_{\alpha\beta}
 = \left(\begin{array}{lll} 0& 0& 0\\ 0 & 0 & 0
\\0 & 0 & 
\eta^{\prime} \end{array}
\right)_{\alpha\beta} + x_\alpha^2 z_\beta^{\prime} 
\eqno(4.9)
$$
However, we can always make rotations on the upper
$2\times 2$ block so that $x^2, z, z^{\prime}$ have 
0 entries in the first component.  Using 
equation (4.9), we easily see that both
{\boldmath$\lambda_D$} and {\boldmath$\lambda_U$} are only non-zero in 
the lower $2\times 2$ block, and CKM mixing only occurs in the
23 sector, as claimed. 
Thus, in order to have, for example, a tree level $\theta_c
$ or $V_{ub}$ (as is necessary from our discussion in section 3),
 we must ensure that 
{\boldmath $\Lambda_D$} and {\boldmath $\Lambda_U$} do not have 
entries in the same
two rows. Other than this case, we expect generically that all elements of the
CKM matrix exist at tree level.

In this section we have shown that if the Higgs couples in only 2 rows
or 2 columns of the full Yukawa matrix to matter, 
then there will be a light
generation which is massless at tree level. The required 
sparseness of Higgs couplings
is due to $G_f$ and holomorphy. 

\section{Realistic models for radiative fermion masses}

\hspace{\parindent} In \cite{ACH1}, we gave explicit lepton models of 
flavor with radiative electron mass, which naturally fulfilled the
phenomenological requirements of Sec. 3;
 namely, the electron is massless at
tree level, the muon picks up a tree level mass upon integrating out heavy
states, muon number is conserved, and   $D$
terms yiels $e - \tau$ mixing which generates
 a radiative electron mass. In this section,
our purpose is to give an extension to the quark sector. We begin by reviewing
the lepton model most readily extended to the quark sector, the full model with
flavor group $G_f = SU(2)_{\ell} \times SU(2)_e \times U(1)_A$. The fields are
categorized as light/heavy and matter/Higgs in Table 2.

\begin{table}[htb]
\begin{tabular}{cc|c} & {\bf Light} & {\bf Heavy}\\
{\bf Matter} & $\ell_3(0),\ell_I(+1)$& $L(+2),L_I(+1),\bar{L}(-2),
\bar{L}^I(-1)$
\\ &$e_3(0),e_i(-1)$& $E(-2),E_i(-1),\bar{E}(+2),\bar{E}^i(+1)$\\
 \hline {\bf Higgs} & $h(0)$ & $\phi_{\ell I}(+1),\phi_{e i}(-1),S(0)$
\end{tabular}
\vskip .3in
{\bf Table 2}: Field content and  $G_f$ transformation properties
for the lepton model. $I, \, i$ are 
$SU(2)_{\ell}$ and $SU(2)_e$ indices respectively, 
the numbers in brackets are the
$U(1)_A$ charges.
\end{table}

We require the theory to be invariant under matter parity ({\it Matter} 
$\rightarrow -${\it Matter}) 
and heavy parity ({\it Heavy} $\rightarrow -${\it Heavy}). 
Here, matter parity is
crucial to avoid dangerous $R$-parity violating couplings, but the heavy parity
is imposed only for simplicity.\footnote{However, both of these parities are
automatic in the $SU(3)_{\ell} \times SU(3)_e$ models considered in \cite{ACH1}.
The $U(1)_A$ factor in $G_f$ also finds a natural explanation in these theories.
We do not use the $SU(3)$ theories here as a starting point here because the
requisite modifications to go to the quark sector are more difficult to see than
in the $SU(2)_{\ell} \times SU(2)_e \times U(1)_A$ model we are considering.}
Requiring these discrete symmetries and $G_f$ invariance gives us the following
renormalizable superpotential (where all dimensionless couplings are $O(1)$)
$$
\eqalignno{
W= &\lambda_3 l_3 e_3 h + \lambda_4 L E h\cr
& +f_1 l_3 \bar{L}^I \phi_{\ell I} +  f_2 \ell_I \bar{L}^I S +  
f_3 \ell_I \epsilon^{IJ} \phi_{\ell J} \bar{L} \cr
& +f_1^{\prime} e_3 \bar{E}^i \phi_{e i} + f_2^{\prime} e_i\bar{E}^i S +
f_3^{\prime} e_i \epsilon^{ij} \phi_{e j} \bar{E}\cr
& +M_L \bar{L} L +M_{L_I} \bar{L}^I L_I + M_E \bar{E} E +
M_{E_i} \bar{E}^i E_i. &(5.1)\cr}
$$
Note that this superpotential has only two Yukawa couplings $\lambda_3$ (for the
$\tau$) and $\lambda_4$ (for the superheavy $L,E$). Therefore, using the results
of the last section, we are guaranteed to have a tree-level massless state after
we integrate out the heavy fields;\footnote{Actually, in this theory the
existence of a massless state can already be seen in the effective theory as
described in \cite{ACH1}.} we identify this state with the electron.

The fields $\phi_{\ell}$, $\phi_e$ and $S$ take VEV's
which break the flavor symmetries. We can assume without
loss of generality that $\VEV{\phi_{\ell}}= (v_{\ell},0), \VEV{\phi_e}
= (v_e,0)$. As described generally in the previous section, these VEV's mix the
light and heavy states and we must rotate to the mass basis where ``light'' and
``heavy'' are properly identified. An approximation to the resulting rotation on
the Yukawa matrix is shown in Fig.\ 5, and we generate the following
superpotential term for the light fields:
$$
\eqalignno{
 \, \Delta W  
& = ({{f_3 \ell_I \epsilon^{IJ} 
\VEV{\phi_{\ell_J}}}\over M_L}) \lambda_4 h 
({{f_3^{\prime} e_i \epsilon^{ij} 
\VEV{\phi_{e_j}}}\over M_E})\cr  &
= \lambda_4 ({{f_3 v_{\ell}}\over M_L})
({{f_3^{\prime} v_e}\over M_E}) \ell_2 e_2 h,& (5.2)\cr}
$$
so, we can identify $(\ell_2,e_2)$ with the muon and $(\ell_1,e_1)$ with the 
electron.

Let us look at the above rotation more directly \cite{DP}. Setting $\phi_{\ell}, 
\phi_e, S$ to their VEV's gives the follwing mass terms in the superpotential:
$$
W_{mass} = M_L \bar{L} (L + \epsilon_{\ell} \ell_2) 
+ M_{L_I} \bar{L}^1 (L_1 + \epsilon_{\ell}^{\prime} \ell_3 + 
\epsilon_{\ell}^{\prime \prime} \ell_1) + M_{L_I} \bar{L}^2 (L_2 + 
\epsilon_{\ell}^{\prime \prime} \ell_2),
\eqno(5.3)
$$
plus similar terms for the $E$'s,
where $\epsilon_{\ell} = -{{f_3 v_{\ell}}\over M_L}, \; 
\epsilon_{\ell}^{\prime}= {{f_1 v_{\ell}}\over M_{L_I}}, \; 
\epsilon_{\ell}^{\prime \prime} = {{f_2 \VEV{S}}\over M_{L_I}}. $
Thus, the mass basis is related to the flavor basis via $\ell ^{\prime} = 
U_{\ell} \ell$, where $\ell^{(\prime) T} = 
(\ell_1, \ell_2, \ell_3, L,L_1,L_2)^{(\prime)}$.
To a first approximation, we have 
$$
U_{\ell} = \left(\begin{array}{cccccc}
1 & 0 & 0 & 0 & - \epsilon_{\ell}^{\prime \prime *}& 0\\
0 & 1 & 0 & -\epsilon_{\ell}^* & 0 & - \epsilon_{\ell}^{\prime \prime *}\\
0 & 0 & 1 & 0 & - \epsilon_{\ell}^{\prime *} & 0\\
0 & \epsilon_{\ell} & 0 &1 & 0 & 0\\
\epsilon_{\ell}^{\prime \prime} & 0 & \epsilon_{\ell}^{\prime} & 0 & 1 &0\\
0 & \epsilon_{\ell}^{\prime \prime} & 0 & 0 & 0 & 1\end{array}\right).
\eqno(5.4)
$$
Completely similar statements hold for the $e$'s. Now, in the original flavor
basis, the Yukawa matrix {\boldmath$\Lambda$} is
$$
\mbox{\boldmath$\Lambda$} = \left(\begin{array}{cccccc}
0 & 0 & 0 & 0 & 0 &0\\ 0 & 0 & 0 & 0 & 0 & 0\\0 & 0 & \lambda_3 & 0 & 0 & 0\\
0 & 0 & 0 & \lambda_4 & 0 & 0\\ 0 & 0 & 0 & 0 & 0 &0\\ 0 & 0 & 0 & 0 & 0 & 0
\end{array}\right).
\eqno(5.5)
$$
After rotating to the mass basis, we have
$$
\mbox{\boldmath$\Lambda$}^{\prime} = U_{\ell}^* \mbox{\boldmath$\Lambda$}
U_e^\dagger = 
\left(\begin{array}{cccccc}
0 & 0 & 0 & 0 & 0 & 0\\0 & \epsilon_{\ell} \epsilon_e \lambda_4 & 0 &
 -\epsilon_{\ell} \lambda_4& 0 & 0\\
0 & 0 & \lambda_3 & 0 & \epsilon_e^{\prime} \lambda_3 & 0\\
0 & - \epsilon_e \lambda_4 & 0 & \lambda_4 & 0 & 0\\
0 & 0 &
\epsilon_{\ell}^{\prime} \lambda_3& 0 & \epsilon_{\ell}^{\prime}
\epsilon_e^{\prime} \lambda_3 & 0\\0 & 0 & 0 & 0 & 0 &0\end{array}\right).
\eqno(5.6)
$$
Dropping all couplings to the heavy states, we obtain the low energy Yukawa
matrix {\boldmath $\lambda$},
$$
\mbox{\boldmath$\lambda$} = \left(\begin{array}{ccc}
0 & 0 & 0\\ 0 & \epsilon_{\ell} \epsilon_{e} \lambda_2 & 0\\ 0 & 0 & \lambda_1
\end{array}\right),
\eqno(5.7)
$$
just as we found earlier.

Note that the VEV's $\VEV{\phi_{\ell}}$ and $\VEV{\phi_e}$ do not completely
break $G_f$; the generator 
$$
T_{\mu} = T_{U(1)_A} - 2(T_{\ell}^3 - T_e ^3)
\eqno(5.8)
$$
annihilates both $\VEV{\phi_{\ell}}$ and $\VEV{\phi_e}$, and corresponds to 
the muon
number:\footnote{ The $U(1)_A$ factor in $G_f$
can be replaced with its $Z_4$ subgroup and still avoid dangerous muon number
violating processes; after the VEV's are taken there is  a symmetry under
$(\ell_2, e_2) \rightarrow (-\ell_2, -e_2)$
 which still forbids mixing between the
scalar $\mu$ and $\tau,e$, therefore avoiding the dangerous $\mu\to e\gamma$
decay.}
$$
e^{i \theta T_{\mu}} \left(\begin{array}{c} \ell_1 \\ \ell_2\end{array}\right)
= \left(\begin{array}{c} \ell_1 \\  e^{2 i \theta} \ell_2\end{array}\right), \;
e^{i \theta T_{\mu}} \left(\begin{array}{c} e_1 \\ e_2\end{array}\right)
= \left(\begin{array}{c} e_1 \\  e^{-2 i \theta} e_2\end{array}\right).
\eqno(5.9)
$$
We now have most of what we want; we need only show that the required mixing
between the $\tau$ and $e$ is generated in the scalar mass matrix. We can
generate $D$ term mixings upon integrating out heavy states \cite{DP}. 
The one in Fig.\ 6 gives
$$
{{f_2 \ell_I \VEV{S}}\over M_{L_I}} {{f_1^* \ell_3^{\dagger}
\VEV{\phi_{\ell 1}}}\over M_{L_I}}
={{f_2 \VEV{S}}\over M_{L_I}} {{f_1^* v_{\ell}^*}\over M_{L_I}}
\ell_1 \ell_3^{\dagger}.
\eqno(5.10)
$$
Note that this term explicitly breaks the $U(1)_{\ell_1}$ chiral symmetry
associated with the zero tree-level Yukawa coupling of the electron, so we
expect the required mixing between
$\widetilde{\tau}$ and $\widetilde{e}$ to occur. Let
us check it more explicitly. The $D$-term part of the
lagrangian is $\int d^4\theta (\phi^{\dagger} \phi + \theta^2 \bar{\theta}^2
\phi^{\dagger} m^2 \phi)$, where $\phi$ is a collection of all the fields and 
$m^2$ is the soft supersymmetry breaking scalar 
mass matrix. When we rotate to the mass
basis, we send $\phi \rightarrow U \phi$. Under this rotation, $\phi^{\dagger}
\phi$ is invariant, but $m^2 \rightarrow U m^2 U^\dagger$.\footnote{This is not
strictly speaking correct, since supersymmetry breaking can affect the rotation
to the mass basis. For instance, in Fig.\ 6, we could attach spurions $\theta^2$
and $\bar{\theta}^2$ to the superpotential vertices, obtaining a direct
contribution to the scalar mass matrix of order $|A|^2$, where $A$ 
is the trilinear
soft term associated with the superpotential vertex. Put another way, we can
have spurions $\theta^2$ in the rotation matrix $U$, and get contributions to
the scalar masses from rotating $\phi^{\dagger} \phi$. These contributions are
of the same order as the ones we are discussing, but do not affect any of our
results.}. 
In our example, the scalar mass term
for the left-handed lepton fields is $\ell^{\dagger} m_{\ell}^2 \ell$, with 
$m_{\ell}^2$ = diag($m_{\ell_I}^2, m_{\ell_I}^2, m_{\ell_3}^2,m_L^2,
m_{L_I}^2, m_{L_I}^2)$.  The scalar mass matrix for the three low energy 
generations is then
$$
\begin{array}{ll}
{\bf m}_{\ell_{\alpha \beta}}^{2 \; (3 \times 3)} &= 
{(U_{\ell} m_{\ell}^2 U_{\ell}^\dagger)}_{\alpha \beta} \\
&= \left(\begin{array}{ccc}
 m_{\ell_3}^2 + |\epsilon_{\ell}^{\prime}|^2 m_{L_I}^2 
& 0 & \epsilon_{\ell}^{\prime *} \epsilon_{\ell}^{\prime \prime} m_{L_I}^2\\
0 & m_{\ell_I}^2 + |\epsilon_{\ell}|^2 m_L^2 + 
|\epsilon_{\ell}^{\prime \prime}|^2 m_{L_I}^2 & 0\\
\epsilon_{\ell}^{\prime} \epsilon_{\ell}^{\prime \prime *} m_{L_I}^2 & 0 &
m_{\ell_I}^2 + |\epsilon_{\ell}^{\prime \prime}|^2 m_{L_I}^2\end{array} 
\right).  \\ 
\\
\end{array}
\eqno (5.11)
$$
The zero entries in the above matrix are a consequence of the unbroken 
$U(1)_{\mu}$ symmetry of the theory.  We can explicitly see the 1-3
entry generated in the scalar mass matrix, 
which, together with the corrseponding 1-3
entry in the the right-handed scalar mass matrix, is responsible for generating
the radiative electron mass.
   
There are two difficulties when we try to extend the lepton model for radiative
electron mass to the quark sector. First, the radiative down quark mass is severely
constrained by $B-\bar{B}$ mixing as we showed in Sec. 3. This can be
resolved if the SUSY-breaking masses are heavy enough ($\gsim$ 1 TeV). The other
problem is that in addition to the quark masses, we also have to get the correct
CKM mixing matrix. As we have shown in Sec. 3, it is very difficult to generate
all CKM mixing matrix elements: squark masses have to be pushed up to 
unacceptably high scales and unnatural flavor mixing gaugino interactions are
needed. 
Excluding that possibility, one has to put in some mixing angles
at tree level. In subsection 5.1 we present a model in which all first
generation fermion masses come from radiative corrections. In subsection 5.2 we
construct a model in which $m_e$ and $m_u$ come from radiative corrections while
$m_d$ and $\theta_c$ appear at tree level with the  prediction $\sin\theta_c =
\sqrt{m_d/m_s}$. We show that this model can be naturally embedded in the flipped 
$SU(5)$ grand unified theory.

\subsection{A complete model for radiative first generation fermion masses}

\hspace{\parindent} The complete model 
for quarks and leptons is based on the same flavor
group $G_f = SU(2)_l \times SU(2)_r \times U(1)_A$ as in 
the lepton model. However,
a minimal direct extension of the lepton model to the quark sector does not give
tree level CKM mixing angles. Following the guidelines to generate tree level
$\theta_c$ and \vub\ in Sec. 4, we need to introduce two heavy left-handed
$SU(2)_l$ singlet quarks $Q$, $Q'$ (and their conjugates $\bar{Q}$, 
$\bar{Q'}$).\footnote{Second 
pairs of heavy $U'$, $\bar{U'}$ and $D'$, $\bar{D'}$ are not
included in our discussion. They can be added as long as their $U(1)_A$ charge
assignments forbid their Yukawa interactions with the $Q$'s and Higgses.} Their
$U(1)_A$ charges are assigned such that $Q$ only couples to the up-type Higgs
but not the down-type Higgs and vice versa for $Q'$. In addition, there cannot
be an unbroken $U(1)$ left in the quark sector, so we introduce a second 
$SU(2)_l$ doublet $\phi'_l$, and a second 
$SU(2)_r$ doublet $\phi'_r$, whose VEV's are in
different directions from the directions of $\phi_l$ and $\phi_r$ VEV's, 
breaking $G_f$ completely. The field content and $G_f$
transformation properties of the quark sector 
are shown in Table 3. We also impose  matter-parity and
heavy parity.
\begin{table}[hbt]
\begin{tabular}{cc|ccc}
& {\bf Light} & \multicolumn{3}{c}{\bf Heavy}\\
& $u_3(0),\,u_i(-1)$ & $U(-2),\,\bar{U}(+2),$ & & $U_i(-1),\,
\bar{U}^i(+1)$ \\
{\bf Matter} & $q_3(0),\,q_I(+1)$ & $Q(+2),\,\bar{Q}(-2),$ & $Q'(0),\,
\bar{Q'}(0),$
& $Q_I(+1),\, \bar{Q}^I(-1)$ \\
& $d_3(0),\, d_i(+1)$ & & $D(0),\, \bar{D}(0),$ & $D_i(+1),\, \bar{D}^i(-1)$\\
\hline {\bf Higgs} & $h_u(0),\, h_d(0)$ & $\phi_{lI}(+1),\, \phi_{ri}(-1),$ &
$\phi'_{lI}(-1),\, \phi'_{ri}(+1),$ & $S(0)$
\end{tabular}
\vskip .3in
{\bf Table 3}: Field content and  $G_f$ transformation
properties of the quark sector. 
$I$ and $i$ are $SU(2)_l$ and $SU(2)_r$ doublet indices and the
numbers in brackets are  $U(1)_A$ charges.
\end{table}
The VEV's of $\phi,\, \phi'$ and $S$ are assumed to take the most general 
form:\footnote{$\phi_{lI},\, \phi_{ri}$ can be put in this form by $SU(2)_l$
and $SU(2)_r$ rotations, then $\phi'_{lI},\, \phi'_{ri}$ VEV's will take the
general directions if there are no alignments between $\phi'_{lI},\,\phi'_{ri}$
 and $\phi_{lI},\, \phi_{ri}$.
In this paper we do not specify the origin of these VEV's.}
$$
\eqalignno{
\vev{\phi_{lI}}&=\left( \begin{array}{c} v_{l0}\\ 0 \end{array} \right),\;
\vev{\phi_{ri}}=\left(\begin{array}{c} v_{r0}\\ 0 \end{array} \right), \cr
\vev{\phi'_{lI}}&=\left(\begin{array}{c} v_{l1}\\ v_{l2}\end{array} \right),\;
\vev{\phi'_{ri}}=\left(\begin{array}{c} v_{r1}\\ v_{r2}\end{array} \right),\;
\vev{S}= v_s . &
(5.12)}
$$
Because we are dealing with a full theory, we restrict ourselves to  
renormalizable interactions only and all possible renormalizable
interactions consistent with the symmetries are included. Nonrenormalizable
interactions are assumed to be absent or suppressed enough so that they can be
ignored. The $G_f$ transformation properties of the up sector are identical
to those of the lepton model so the analysis is exactly the same as in the 
lepton model. The superpotential for the up sector is
$$
\eqalignno{
W_u = & \lambda_{u3} q_3 h_u u_3 + \lambda_{u4} Q h_u U \cr
&+ f_{q1} q_3 \bar{Q}^I \phi_{lI} + f_{q2} q_I \bar{Q}^I S 
+ f_{q3} \epsilon^{IJ} q_I \bar{Q} \phi_{lJ} \cr
&+ f_{u1} u_3 \bar{U}^i \phi_{ri} + f_{u2} u_i \bar{U}^i S 
+ f_{u3} \epsilon^{ij} u_i \bar{U} \phi_{rj} \cr
&+ M_U \bar{U}U + M_{U_i} \bar{U}^i U_i + M_Q \bar{Q}Q+M_{Q_I}\bar{Q}^I Q_I .
&(5.13)}
$$
Note that although we introduce another pair of $G_f$ breaking fields
$\phi'_{lI}$ and $\phi'_{ri}$, they do not have renormalizable
interactions with the up sector and the lepton sector. The only
such $G_f$ invariant interactions
$$
L \bar{L}^I \phi'_{lI},\; E \bar{E}^i \phi'_{ri},\; Q \bar{Q}^I \phi'_{lI},\;
U \bar{U}^i \phi'_{ri}
\eqno{(5.14)}
$$
are forbidden by heavy parity. Therefore, we do not generate
muon number violating operators even though $G_f$ is completely broken.

The superpotential of the down sector is given by
$$
\eqalignno{
W_d = & \lambda_{d3} q_3 h_d d_3 + \lambda_{d4} Q' h_d D \cr
&+ f'_{q3} \epsilon^{IJ} q_I \bar{Q'} \phi'_{lJ} +f'_{q4} q_3 \bar{Q'} S\cr
&+ f_{d1} d_3 \bar{D}^i \phi'_{ri} + f_{d2} d_i \bar{D}^i S 
+ f_{d3} \epsilon^{ij} d_i \bar{D} \phi_{rj} +f_{d4} d_3 \bar{D} S\cr
&+ M_D \bar{D}D + M_{D_i} \bar{D}^i D_i + M_{Q'} \bar{Q'}Q' .
& (5.15)}
$$
The $f_{d1}$ and $f_{d2}$ couplings are responsible for the $D$-term mixing
between $d_3$ and $d_i,\, i=1,2$ (with intermediate $\bar{D^i}$).
$f_{d3},\, f_{d4}$ mix $d_2,\, d_3$ with $D$, $f'_{q3},\,f'_{q4}$ mix
$q_1,\, q_2,\, q_3$ with $Q'$ and they are responsible for generating tree 
level Yukawa couplings among $d_2,\, d_3,$ and $q_1,\,q_2,\,q_3$ with $h_d$. 
After integrating out the heavy states, we obtain the following tree level
Yukawa matrices for the up quarks and down quarks:
$$
\mbox{\boldmath $\lambda$}_U =\left(
\begin{array}{ccc} 0& 0& 0 \\ 0& \epsilon_{q2}\epsilon_{u2}\lambda_{u4} & 0 \\
0 & 0& \lambda_{u3}
\end{array} \right),\,
\mbox{\boldmath $\lambda$}_D = \left(
\begin{array}{ccc}
0& \epsilon'_{q1}\epsilon_{d2}\lambda_{d4} 
& \epsilon'_{q1}\epsilon_{d3}\lambda_{d4} \\
0& \epsilon'_{q2}\epsilon_{d2}\lambda_{d4} 
& \epsilon'_{q2}\epsilon_{d3}\lambda_{d4} \\
0& \epsilon'_{q3}\epsilon_{d2}\lambda_{d4} 
& \epsilon'_{q3}\epsilon_{d3}\lambda_{d4} + \lambda_{d3}
\end{array} \right),
\eqno{(5.16)}
$$
where,
$$
\eqalignno{
\epsilon_{q2} &={f_{q3} v_{l0}\over M_Q},\;
\epsilon_{u2}={f_{u3} v_{r0}\over M_U},\cr
\epsilon'_{q1} &=-{f'_{q3} v_{l2}\over M_{Q'}},\;
\epsilon'_{q2}={f'_{q3} v_{l1}\over M_{Q'}},\;
\epsilon'_{q3}={f'_{q4} v_{s}\over M_{Q'}},\cr
\epsilon_{d2} &={f_{d3} v_{r0}\over M_D},\;
\epsilon_{d3}={f_{d4} v_{s}\over M_D}.}
$$
Both matrices are of rank 2, as guaranteed by the theorem 
of Sec.\ 4, (although
this  cannot be seen from the effective theory point of view). Now we have a 
massless state in each of the up and down sectors and all mixing angles are 
generated at tree level. $m_u$ and $m_d$ are then generated radiatively
by the mixings between the first and the third generations induced by $f_{q1},
\,f_{q2},\,f_{u1},\,f_{u2},$ and $f_{d1},\,f_{d2}$ with intermediate
$\bar{Q'},\, \bar{U},$ and $\bar{D}$ states. $f_{d3},\,f_{d4},\,f'_{q3},\,
f'_{q4}$ also induce the $D$ term mixings among generations with intermediate
$\bar{D}$ and $\bar{Q'}$ states. For example, the mixing between $q_3$ and 
$q_2$ is $\sim \epsilon'_{q3} \epsilon'_{q2}$, which is about the same size
as the corresponding CKM mixing angle. For large \tb\,  they can give sizable
corrections [${\cal O}(50\%)$] to the CKM matrix elements. Since we
do not know the 
exact size and the sign of these corrections, if we just take $m_s$, \sc\,
and \vcb\ to be approximately equal to the tree level results, then we have
[within ${\cal O}(50\%)$ accuracy]
$$
\eqalignno{
V_{cb} & \simeq \;\epsilon'_{q2}\epsilon_{d3} {\lambda_{d4} \over \lambda_{d3}}
\; \simeq \; 4\times 10^{-2}, \cr
{m_s\over m_b} & \simeq \;\epsilon'_{q2}\epsilon_{d2} 
{\lambda_{d4} \over \lambda_{d3}} \;\simeq \; 2.7\times 10^{-2}, \cr
\sin\theta_c & \simeq \;{\epsilon'_{q1}\epsilon_{d2}\over \epsilon'_{q2}
\epsilon_{d2}} \hskip .25in \simeq \; 0.22 
& (5.17)}
$$
Combining the above relations, we obtain the approximate tree level \vub\,
$$
V_{ub}^{\mbox{\scriptsize{tree}}}\; \simeq \; \epsilon'_{q1}\epsilon_{d3}
{\lambda_{d4} \over \lambda_{d3}}\; \simeq \; \sin\theta_c V_{cb}\;
\simeq \;9\times 10^{-3},
\eqno{(5.18)}
$$ 
which is about a factor of 2  bigger than the accepted value. However, as 
we found in Sec.\ 3, when we generate $m_d$ by radiative corrections, we also
generate $V_{ub}^{\mbox{\scriptsize{rad}}}$ bigger than the accepted value
by about a factor of 3, which has to be cancelled by the tree level
$V_{ub}^{\mbox{\scriptsize{tree}}}$. If the sign is right, (5.18) 
is just in the 
range which can cancel against the radiative contribution to produce the 
correct \vub . Therefore, correct values for all quark masses and CKM mixing
angles can be obtained. Naively, one might expect that it is difficult
to have massless first generation quarks at tree level because of the Cabibbo
angle. Here we showed, with the help of the theorem of Sec.\ 4 for the rank
of the Yukawa matrices, that one can naturally get massless up and down
quarks at tree level, while having nonzero \sc .

\subsection{A model of radiative $m_u$, $m_e$, and tree level $m_d$}

\hspace{\parindent} As we have mentioned, 
a radiative $m_d$ is only barely consistent
with $B-\bar{B}$ mixing with very heavy SUSY-breaking masses. In this
subsection, we present a model in which $m_d$ is nonzero at tree level, while
$m_u$ and $m_e$ arise purely from radiative effects. The flavor group is
$G_f=SU(2)_T\times SU(2)_F \times Z_4$. The reason for the subscripts of the
$SU(2)$ groups will be clear later. $U(1)_A$ is replace by its subgroup
$Z_4$.  Matter-parity and the heavy parity are imposed as well. The field content
is shown in Table 4, where $I$, $i$ are $SU(2)_F$ and $SU(2)_T$ indices
respectively, and the numbers in brackets are the $Z_4$ charges with
$n$ and ($n \bmod 4$) identified.
$\phi_{Ti},\, \phi_{FI},\, S$ and $X$ have nonzero VEV's:
$$
\vev{\phi_{Ti}} =\left(\begin{array}{c} v_T \\ 0 \end{array} \right),\;
\vev{\phi_{FI}} =\left(\begin{array}{c} v_F \\ 0 \end{array} \right),\;
\vev{S}=v_s,\; \vev{X}=v_x, 
\eqno{(5.19)}
$$
which break $G_f$ completely.
\begin{table}[htb]
\begin{tabular}{cc|ccc}
& {\bf Light} & \multicolumn{3}{c}{\bf Heavy}\\
& $e_3(0),\, e_i(-1)$ & $E(-2),\, \bar{E}(+2)$ & $E_i(-1),\, \bar{E}^i(+1)$ & \\
& $\ell_3(0),\, \ell_I(+1)$ & $L(+2),\, \bar{L}(-2)$ & $L_I(+1),\, 
\bar{L}^I(-1)$ & \\
{\bf Matter}& $u_3(0),\,u_I(+1)$ & $U(+2),\,\bar{U}(-2),$ &  $U_I(+1),\,
\bar{U}^I(-1)$ &\\
& $q_3(0),\,q_i(-1)$ & $Q(-2),\,\bar{Q}(+2),$ & $Q_i(-1),\, \bar{Q}^i(+1),$ 
& $Q'_i(+1),\, \bar{Q'}^i(-1)$ \\
& $d_3(0),\, d_i(-1)$ & $D(-2),\, \bar{D}(+2),$ & $D_i(-1),\, \bar{D}^i(+1),$
& $D'_i(+1),\, \bar{D'}^i(-1)$ \\
\hline {\bf Higgs} & $h_u(0),\, h_d(0)$ & $\phi_{Ti}(-1),\, \phi_{FI}(+1),$ &
$S(0),\; X(2)$ &
\end{tabular}
\vskip .3in
{\bf Table 4}: Field content and $G_f$ transformation properties of the 
model with radiative $m_u$, $m_e$, and tree level $m_d$.
\end{table}
 In this model there is only one pair
of $SU(2)_{T,F}$ breaking fields $\phi_{Ti},\,\phi_{FI}$. The tree level 
massless electron and up quark can be easily seen in an effective theory
point of view\cite{ACH1}, because the only $SU(2)_{T,F}$ invariant holomorphic
combinations of the two light generations and fields with nonzero VEV's for
the lepton and the up quark sectors are $\epsilon^{ij} e_i \phi_{Tj}$,
$\epsilon^{IJ} \ell_I \phi_{FJ}$, $\epsilon^{IJ} u_I \phi_{FJ}$, and 
$\epsilon^{ij} q_i \phi_{Tj}$, which cannot give Yukawa couplings to both
light generations with $h_u$ and $h_d$. In the down sector, $q$'s and $d$'s
have the same $G_f$ transformation properties. One can write down the 
effective operator
$$
\epsilon^{ij} q_i h_d d_j X S ,
\eqno{(5.20)}
$$
which generates the 12 and 21 entries of the down Yukawa matrix with equal
size and opposite signs. Hence we can obtain both $\theta_c$ and $m_d$ at
tree level with the experimentally successful relation $\sin \theta_c
\simeq \sqrt{m_d/m_s}$.

Compared with the lepton model discussed earlier in this section, the extra
$X$ field is required to break the left over ``second generation parity''
in order to generate \vcb\ and \vus\, but it may also induce a too big 
$\mu\to e\gamma$ rate, which will be discussed later. The $Q'_i,\, 
\bar{Q'}^i,\, D'_{i},\, \bar{D'}^i$ are responsible for generating the
operator (5.20). They can be omitted if nonrenormalizable operators are
allowed and are sufficiently large. In fact, because this model can be analyzed
in the effective theory point of view, including nonrenormalizable interactions
will not affect our results. However, for simplicity and completeness,
we will analyze the full theory and restrict ourselves to renormalizable
interactions.

The lepton sector and the up quark sector are similar to the previous models.
We will not repeat the detailed analysis. The only difference is that with
the additional $X$ field, we can have the following extra interactions:
$$
f_{e5}X e_3 \bar{E},\, f_{\ell 5}X \ell_3 \bar{L},\, f_{u5}X u_3 \bar{U},\,
f_{q5} X q_3 \bar{Q} .
\eqno{(5.21)}
$$
They mix the third generation with the heavy $SU(2)_{T(F)}$ singlet generation.
In combination with $\epsilon^{ij} \phi_{Ti} e_j \bar{E}$, $\epsilon^{IJ}
\phi_{FI} \ell_J \bar{L}$, $\epsilon^{IJ} \phi_{FI} u_J \bar{U}$,
and $\epsilon^{ij} \phi_{Ti} q_j \bar{Q}$, they generate the 23 and 32
entries of the Yukawa matrices and also the $D$ term mixing between the 
second and the third generations. For the up quark sector, the $D-\bar{D}$
mixing constraints are very weak and hence easily satisfied. However, for
the lepton sector the constraint from the $\mu\to e\gamma$ rate requires the
2-3 mixing to be no bigger than ${\cal O}(10^{-3})$, while the naive 
expectation of 2-3 mixing in this model is of the order \vcb .
Therefore, one has to assume that the couplings of the $X$ field to the
lepton sector are small, or prevented by some extra symmetries. We will
see that this is possible to  achieve later. 

In the down quark sector,
in addition to the usual interactions,
$$
\eqalignno{
W_d = & \lambda_{d3} q_3 h_d d_3 + \lambda_{d4} Q h_d D \cr
&+ f_{q1} q_3 \bar{Q}^i \phi_{Ti} + f_{q2} q_i \bar{Q}^i S 
+ f_{q3} \epsilon^{ij} q_i \bar{Q} \phi_{Tj} \cr
&+ f_{d1} d_3 \bar{D}^i \phi_{Ti} + f_{d2} d_i \bar{D}^i S 
+ f_{d3} \epsilon^{ij} d_i \bar{D} \phi_{Tj} \cr
&+ M_D \bar{D}D + M_{D_i} \bar{D}^i D_i + M_{Q}\bar{Q}Q+ M_{Q_i}\bar{Q}^i Q_i ,
& (5.22)}
$$
which give the tree level $b$ and $s$ quark masses and 1-3 $D$ term mixing,
we have the following interactions as well,
$$
\eqalignno{
W'_d = & f_{q5} q_3 \bar{Q} X + f_{d5} d_3 \bar{D} X \cr
&+ f_{q6} q_i \bar{Q'}^i + f_{d6} d_i \bar{D'}^i X \cr
&+ \lambda_{d5} \epsilon^{ij} Q'_i h_d D_j + \lambda_{d6}\epsilon^{ij}
Q_i h_d D'_j .
& (5.23)}
$$
As we have discussed before, the $f_{q5},\, f_{d5}$ couplings induce the 
23 and 32 entries of the Yukawa matrix and the 2-3 $D$ term mixing, so that
\vcb\ can be generated. $f_{q6},\, f_{d6},\, \lambda_{d5},\, \lambda_{d6}$
together with $f_{q2},\, f_{d2}$ couplings generate the operator (5.20),
which gives $\theta_c$ and $m_d$, and the successful relation $\sin \theta_c
= \sqrt{m_d/m_s}$. The tree level down quark mass matrix takes the 
following form,
$$
\left( \begin{array}{ccc} 0 & C & 0 \\ -C & E & B \\ 
0 & B^{\prime} & A \end{array}
\right),
\eqno{(5.24)}
$$
while the tree level up quark and lepton mass matrices have nonzero entries
in the lower $2\times 2$ block. In addition to $m_u$ and $m_e$, \vub\ is 
also generated by radiative corrections from the 3-1 mixing $\wdl_{31}$.
The required size of $\wdl_{31}$ is much smaller than that required for 
generating $m_d$ radiatively, so the phenomenological constraints are easier
to satisfy as we have discussed in \mbox{Sec.\ 3}.

Looking at the $G_f$ transformation properties of the fields, one
can see that this model can be embedded into the flipped $SU(5)$ grand
unified theory\cite{Barr}: $q$ and $d$ (and the not discussed right-handed 
neutrino $n$) belong to the {\bf 10} representation of flipped $SU(5)$,
$u$ and $\ell$ belong to the ${\bf \bar{5}}$ and $e$ is a singlet {\bf 1}
under flipped $SU(5)$. $SU(2)_T$ is a flavor group for the {\bf 10}'s and
$SU(2)_F$ is a flavor group for the ${\bf \bar{5}}$'s. In Table 4, the 
$e$'s are assigned to transform under $SU(2)_T$. Here one can either have them
transform under a different $SU(2)_S$, or simply identify $SU(2)_S$ with
$SU(2)_T$.

One nice feature of embedding this model into flipped $SU(5)$ is that the 
$X$ field can be assigned to the {\bf 75} of $SU(5)$. Because only the
${\bf 10}\times {\bf \overline{10}}$ contains {\bf 75} and the ${\bf 5}\times
{\bf \bar{5}},\; {\bf 1}\times {\bf 1}$ do not, the $X$ field can only couple
to $q$ and $d$ but not the lepton sector. Then the $\mu$-$\tau$ mixing and 
hence the troublesome $\mu\to e\gamma$ decay rate can be removed.

After flipped $SU(5)$ is broken, we do not expect the couplings and the 
mixings to be the same for fields belonging to the same representations of 
the flipped $SU(5)$.\footnote{If flipped $SU(5)$ were not broken, the tree 
level 12 and 21 entries of the down quark mass matrix would not be generated, 
because $\epsilon^{ij} {\bf 10}_i {\bf 10}_j h_d X S$ vanishes. However,
since the flipped $SU(5)$ is broken, $q$'s and $d$'s can have different
mixings so that $\epsilon^{ij} q_i d_j h_d X S$ can be nonzero.}
But if we assume that they are of the same order, the radiative $m_e$, $m_u$
and \vub\ are also consistent: radiative \vub\ does not need a big $\wdl_{31}$
($\sim 10^{-2}$), then $\wur_{31}$ has to be quite big ($\gsim 10^{-1}$) for
generating $m_u$; but so is its flipped $SU(5)$ 
partner $\wel_{31}$ for generating
$m_e$. On the other hand, $\lambda_U$, $\lambda_D$, and $\lambda_E$ are
independent in flipped $SU(5)$ models. They can take suitable values so that
all the tree level quantities come out correctly.

It is possible to extend the $SU(2)$ flavor groups to $SU(3)$ for these
quark models as we did for the lepton model in \cite{ACH1}.
However, here we do not gain much by paying the price that the third
generation Yukawa couplings arise at the nonrenormalizable level.
More heavy fields have to be introduced and more complicated stages of
flavor symmetry breakings are involved. Therefore, we will not pursue
this direction further in this paper.

\section{Conclusions}

\hspace{\parindent} In this paper, we have considered the possibility
of generating some of the light fermion masses through radiative corrections.
Any theory of radiative fermion masses must have an accidental symmetry
for the Yukawa sector guaranteeing the absence tree level masses, while
this symmetry must be broken elsewhere in the theory for any mass to be
generated radiatively. In our discussion, supersymmetry has been crucial
in naturally implementing this scenario:  supersymmetric theories
automatically have two sectors (the superpotential and $D$ terms) which need
not have the same symmetries; because of holomorphy the superpotential
may have accidental symmetries not shared by the $D$ terms. Furthermore, the
particles in the radiative loop generating the fermion masses are just 
the superpartners of known particles, and must be near the weak scale 
if supersymmetry is to solve the hierarchy problem. Thus,  supersymmetric 
theories of radiative fermion masses can lead to testable predictions. Working with
supersymmetric theories with minimal low energy field content, we found (with
the plausible assumption that the accidental flavor symmetries 
of the tree level Yukawa matrix are only broken by soft scalar masses)
that FCNC constraints allow only the first generation fermion masses
to have a radiative origin.  

In the lepton sector, a rather large mixing
between the selectron and stau is needed in order to generate the electron
mass. This implies that mixing with the smuon must be highly 
suppressed in order to avoid too large a rate for $\mu \rightarrow 
e \gamma$. The large selectron-stau  mixing also  gives rise to a significant 
rate for $\tau \rightarrow e \gamma$ which is only a factor 10-100
lower than the current experimental limit. 

In the quark sector, in addition to the quark masses, 
the CKM mixing matrix must also be obtained. The FCNC constraints strongly 
limit the possibilities of generating light quark masses and mixing
angles. We found that $m_u$ and \vub\ can be generated by radiative
corrections, while radiatively generating any of $m_d$, $\theta_c$,
and \vcb\ requires heavy scalar masses ($\sim$ 1TeV). Further, it is
very difficult to generate $m_d$, $\theta_c$, and \vub\ together radiatively
unless the scalar masses are between 2 and 20 TeV,
which we view as unacceptably high. These constraints cause the 
principle difficulties in constructing a model of quark flavor with 
radiative masses.

We extended the lepton model with flavor group $SU(2)_\ell \times
SU(2)_e \times U(1)_A$ in \cite{ACH1} to the quark sector. The lepton
model has a number of nice features: the $SU(2)$ breaking $\phi$ 
VEV's are responsible for both $D$-term mixing between the first and
the third generation and generation of the second generation mass,
so the ratio between the radiatively generated first generation mass
and the second generation mass is naturally of the order $1/(16\pi^2)$.
Further, muon number is conserved so that the dangerous rate for 
$\mu \to e\gamma$ is avoided.  A direct extension of 
this model to the quark sector cannot generate the correct CKM mixings,
which requires the addition of 
 more fields and flavor symmetry breakings to the 
theory. 

We presented two complete models with radiative fermion masses.
In the first model, all first generation fermion masses come from
radiative corrections, and there are also  tree level contributions to
$\theta_c$ and \vub\ as required by the FCNC constraints. First 
generation fermions are  guaranteed to be massless
at tree level by requiring the ``big'' Yukawa 
matrices of the full theory to be rank 2. Requiring a tree level
$\theta_c$ and \vub\ forces us to add another heavy left-handed quark $Q'$
and its conjugate $\bar{Q'}$, and another pair of $SU(2)_{l,r}$ flavor
symmetry breaking fields $\phi'_{l,r}$. Muon number is still conserved as
a consequence of the field content and charge assignments of the theory.
With these minimal extensions, we obtain a complete theory of radiative
first generation fermion masses with successful values for CKM mixing
angles.

In view of the fact that a radiative $m_d$ and $B-\bar{B}$ mixing
are only compatible for very heavy scalar masses, we also constructed
a second model in which $m_u$ and $m_e$ come from radiative corrections
but $m_d$ and $\theta_c$ arise at tree level with the successful relation
$\sin\theta_c =\sqrt{m_d/m_s}$. The dangerous $\mu\to e\gamma$ rate can be 
naturally suppressed if we embed this model into the flipped $SU(5)$ 
grand unified theory.

\section*{Acknowledgements}

\hspace{\parindent} 
LJH thanks Scott Thomas for valuable conversations about electric dipole
moments in theories with radiative fermion masses.  
This work was supported in part by the Director, Office of Energy
Research, Office of High Energy and Nuclear Physics, Division of High Energy
Physics of the U.S. Department of Energy under contract No. DE-AC03-76SF00098
and in part by the National Science Foundation under Grant No. PHY-90-21139.
The work of N.A-H is supported by  NSERC.
  
\section*{Appendix A}
\hspace{\parindent} In this appendix, 
we consider the possibility that the soft 
supersymmetry breaking trilinear $A$ terms do not respect the chiral
symmetries of the Yukawa matrix \cite{wyl,banks}.
Before beginning the discussion of radiative fermion masses in this
scenario, let us consider the constraints imposed 
on the form of the $A$ matrix by requiring the desired vacuum
to be the global minimum of the potential. (The extent to which this 
is a neccesity is discussed at the end of this appendix).
Consider  the lepton sector for simplicity 
(identical arguments hold for the quark sector).
Let us work in a basis where the lepton Yukawa matrix is diagonal 
and has $K$ zeros. There are $D$-flat directions in field space where
the right and left handed lepton fields and the down type Higgs
are nonzero. If we restrict ourselves to the $K$ massless 
generations, there are no quartic terms in the potential along
the $D$-flat directions; all we have are the cubic $A$ terms and
the scalar masses. But, if the $A$ terms are non-zero in the
$K \times K$ block of the massless generations, there will
be directions in field space where the cubic terms become
indefinitely negative and cannot be stabilized by the 
quadratic mass terms.
This can only be  avoided  if the 
 $A$ terms are zero in the $K\times K$
block of the $K$ massless generations. This constraint is in 
itself quite powerful. For instance, if $K=3$, we must
have that the $A$ matrix is zero, and the argument that 
one cannot generate any radiative masses goes through exactly
as in section 2. Next, let us consider the case $K=2$. In this
case, the $A$ matrix must be zero in the upper $2\times 2$ block.
Note that we can make a rotation on the first two generation 
scalars to make $A_{i 3}, A_{3 i}$ zero for either $i=1$ or $i=2$.
Now, the potential is no longer unbounded below, but there is
still a local minimum along the $D$-flat directions for the first
two generations where both left and right handed fields aquire
VEV's, breaking electric charge. We  require that the energy
of this minimum is greater than that of the usual minimum, which is
$-{1\over 4} M^2_Z v^2$. For scalars much heavier than $(M_Z v)^{
{1\over 2}}$ = 150 GeV, we can approximate this requirement by
 demanding  that the electric charge breaking minimum has energy 
greater than zero. A straightforward calculation analogous to that
in \cite{FJR} then gives us the following constraint, where we 
assume that all relevant scalars are degenerate with mass $m$:
$$
{1\over 3} (|A_{33}| + |A_{3i}| + |A_{i3}|) \lsim \lambda_3 m .
\eqno(A.1)
$$
There are corrections to this inequality
due to the fact that the true vacuum energy is not zero
but $-{1\over 4} M^2_Z v^2$; still assuming  $m \gsim$ 150 GeV
the correction takes the form:
$$
{1\over 3} (|A_{33}| + |A_{3i}| + |A_{i3}|) \lsim \lambda_3 m ( 
1+ {1\over {2 \lambda_3}} {{M_Z v}\over {m^2}}).
\eqno(A.2)
$$

With these constraints in hand, we begin the phenomenological
analysis.
Suppose that the scalar masses did not break the chiral symmetries
of the Yukawa matrix. Then, since one of $A_{31,13},A_{32,23}$ can
be chosen to be zero by rotations,
one generation would remain 
massless to all orders of perturbation theory. Thus, in order
to generate both generations radiatively, we must have that both
the $A$ terms and the scalar masses break the chiral symmetries
of the Yukawa sector. In the following, we consider the possibility 
that the $A$ terms generate one mass radiatively while the scalar
masses generate the other mass. It is easy to see that this is 
impossible in the lepton sector: the muon mass is too big to be
generated radiatively, and even if we could, we would generate
too large a rate for $\tau \to \mu \gamma$. Moving on to the
quark sector, we have four cases to consider:

(1) $m_d$ from scalar masses and $m_s$ from $A$ terms: In the mass 
insertion approximation, assuming for simplicity that all scalars are
degenerate with mass $m$, we have in the large tan$\beta$ limit
$$
{{m_s}\over {m_t}} = {{\alpha_s}\over{18 \pi}}({{\mu M_{\tilde{g}}}\over {m^2}})
{{(A^d_{23} v_d)}\over m^2} {{(A^d_{32} v_d)}\over m^2}.
\eqno(A.3)
$$
From equation (A.1), however, we must have that 
${{(A^d_{23,32} v_d)}\over m^2} \lsim {{m_b}\over m}$, so
$$
{{m_s}\over {m_t}} \lsim 2\times 10^{-3}   
({{\mu M_{\tilde{g}}}\over {m^2}}) 
{{m^2_b}\over {m^2}}
\eqno(A.4)
$$
which, even for $m$=100GeV, gives too small a value for 
$m_s$ by a factor of $\sim$ 100.

(2) $m_d$ from $A$ terms and $m_s$ from scalar masses:
The same argument as in case (1) suggests that the
generated mass for $m_d$ will be too small by a 
factor of $\sim$ 10. Perhaps this factor can be overcome
for some choice of parameters. However, the scalars are
so light that the required mixing in the scalar mass
matrix to generate $m_s$, together with the $A$ terms
responsible for $m_d$, give unacceptable contributions to
$K_1 - K_2$ mixing, and, if there are $CP$ violating phases,
even more unacceptable contributions to $\epsilon$.

(3) $m_u$ from scalar masses and $m_c$ from $A$ terms:
The general problem with the up sector is that $m_c$
seems to be too heavy to be radiative. In the case we
are considering, we find analogously to equation (A.4)
$$
{{m_c}\over {m_t}} \lsim 2\times 10^{-3} 
({M_{\tilde{g}}\over m})  
({{m_t}\over m})^2
\eqno(A.5)
$$
and so to generate large enough $m_c$ we must again have
fairly light squarks.

(4) $m_u$ from $A$ terms and $m_c$ from scalar masses:
In this case again it is difficult to get a large enough
mass for the charm. In analogy to equation (3.10) we have,
(in the limit where we decouple the first two generations,
minimizing the super-GIM cancellation and so maximizing the
generated charm mass)  
$$
{{m_c}\over {m_t}} = {{ 2 \alpha_s}\over {3 \pi}} 
{{A^u_{33}} \over {M_{\tilde{g}}}} \times
\wul_{31} \wur_{31} \wul_{33}^* \wur_{33}^* 
I({{m^2_{\tilde t}}\over {M^2_{\tilde{g}}}}).
\eqno(A.6)
$$
The maximium value of $\wul_{31} \wur_{31} \wul_{33}^* \wur_{33}^*$
consistent with the unitarity of the $W$ matrices is ${1\over 4}$. Then, we have
$$
{{m_c}\over {m_t}} \lsim 5 \times 10^{-3}
{{A^u_{33}}\over {M_{\tilde{g}}}}
I({{m^2_{\tilde t}}\over {M^2_{\tilde{g}}}}).
\eqno(A.7)
$$
Recalling that $I(1) = {1\over 2}$, we see that, even with
maximal mixing angles, the radiative charm mass is too small
or perhaps right on the edge. However, having such large mixing
in the left handed up 32 sector also implies large mixing in
the left handed down 32 sector, which violates the bounds from
$b \rightarrow s \gamma$ unless the third generation 
scalars are pushed above 
1 TeV. This then makes it difficult to generate a large enough
up mass, since the $A$ term contribution is suppressed by 
$({{m_t}\over m})^2$ from (A.1). We find
$$
{{m_u}\over {m_t}} \lsim 2\times 10^{-5} 
({M_{\tilde{g}}\over m}) (1 + {{\mu \cot \beta}\over {A_{33}}}) 
({{1.7 \mbox{TeV}} \over m})^2
\eqno (A.8)
$$
which is also on the edge. Another difficulty with having
such large 32 mixing is that it  disturbs the degeneracy
between the scalar masses of the first two 
generations for  both left handed up and down squarks, and this
could again give problems with $K_1 - K_2$ mixing and $\epsilon$. 

The above arguments certainly do not rule out the possibility
of generating both light generations radiatively; there may 
be regions of parameter space where our rough bounds are evaded.
 Indeed, it may even be the case that requiring the desired vacuum to have 
lower energy than the charge breaking minima is not necessary, 
perhaps the lifetime of the false vacuum can be long enough for the universe
to have stayed in it up to the present; this remains to be seen.
 However, these arguments, together with the
fact that for  the $A$ terms not to share the same chiral symmetries as
the Yukawa matrices we must  entangle flavor symmetry breaking and 
supersymmetry breaking, provide us with sufficient motivation to restrict
our detailed treatment to the scenario considered in this paper.
  
\section*{Appendix B}

\hspace{\parindent} In \cite{HKT,GMS}, the SUSY FCNC constraints 
are expressed in terms of the
ratios of the off-diagonal scalar masses and the ``universal squark or
slepton masses''. For example, the supersymmetric contribution to the 
$B-\bar{B}$ mixing is given by:\footnote{We use the notation and the formula
in \cite{HKT}, corrected by \cite{GMS}}
$$
\eqalignno{
\Delta M_B^{SUSY} =& {\alpha_s^2\over 216 M_{\tilde{q}}^2} {2\over 3}
f_B^2 m_B \{ (\delta^d_{13})_{LL}^2 [-66\tilde{f}_6(x) -24 x f_6(x)] \cr
&+ (\delta^d_{13})_{RR}^2 [-66\tilde{f}_6(x) -24 x f_6(x)] \cr
&+ (\delta^d_{13})_{LL} (\delta^d_{13})_{RR} [-12 \tilde{f}_6(x)
-456 x f_6(x)] \cr
&+ (\delta^d_{13})_{LR}^2 [132 x f_6(x)] 
 + (\delta^d_{13})_{RL}^2 [132 x f_6(x)] \cr
&+ (\delta^d_{13})_{LR} (\delta^d_{13})_{RL} [228 \tilde{f}_6(x) \},
& (B.1)}
$$
where,
$$
\eqalignno{
f_6(x) &= {1\over 6 (1-x)^5} (-6\ln x -18 x\ln x -x^3 +9 x^2 +9x-17),\cr
\tilde{f}_6(x) &= {1\over 3(1-x)^5} (-6x^2\ln x -6x\ln x +x^3 +9 x^2 -9x-1)
&(B.2)}
$$
are the Feynman loop intrgrals defined in \cite{HKT}, and
$$
x={M_{\tilde{g}}^2\over M_{\tilde{q}}^2},\; 
(\delta^d_{ij})_{LL}= {\delta \tilde{m}^2_{\bar{d}_L b_L} \over 
M^2_{\tilde{q}}}, \;\mbox{and so on.} 
$$
Demanding that each term is no bigger than the experimental 
value of $\Delta M_B$ gives the constraints on $\delta^d_{ij}$.
However, with large splitting in scalar masses of the first two and the
third generations, it is better to have constraints directly on the 
mixing matrix elements because of the ambiguity of what $M_{\tilde{q}}$
should be. In this appendix, we will convert the constraints on 
$\delta_{ij}$ into constraints on the mixing matrix elements $W_{ij}$ directly.

We assume degeneracy between the left-handed and the right-handed 
scalar masses, and also the first two generation scalar masses (denoted
by $m_1$). To reduce the number of parameters, we also assume that the
relevant gaugino mass is degenerate with the third generation scalar mass
(denoted by $m_3$). We also take the
 chirality-changing scalar masses much smaller than the
chirality-conserving ones, so that the eigenstates and eigenvalues are not
disturbed significantly. Now we can express the SUSY FCNC contributions
by the mixing matrix elements and the two parameters $m_3$ and $y\equiv
{m_1^2\over m_3^2}$. For example, the first term in (B.1) becomes
$$
\eqalignno{
& {\alpha_s^2\over 216 M_{\tilde{q}}^2} {2\over 3} f_B^2 m_B
\left({\wdl_{31} (m_1^2-m_3^2)\over m_3^2}\right)^2
[-66\tilde{f}_6(y) + 24 f_6(y)] \cr
=& {\alpha_s^2\over 216 M_{\tilde{q}}^2} {2\over 3} f_B^2 m_B
(\wdl_{31})^2 (y-1)^2
[-66\tilde{f}_6(y) + 24 f_6(y)]. &(B.3) }
$$
Demanding it to be smaller the $\Delta M_B^{\mbox{\scriptsize{EXP}}}$ gives 
the constraint on 
$\wdl_{31}$,
$$
\sqrt{\mbox{Re}|\wdl_{31}^2|} < {18 m_3\over \alpha_s f_B}
\sqrt{{\Delta M_B \over m_B}} (y-1)^{-1} 
[-66\tilde{f}_6(y) + 24 f_6(y)]^{-{1\over 2}} . \eqno{(B.4)}
$$
Similarly, we can obtain constraints on other mixing matrix elements from
the other terms. The constraints from $B-\bar{B}$ mixing are shown in
Table B1(a).

For $K-\bar{K}$ mixing, $\Delta {m^2_{LL(RR)}}_{21}$ can have two 
contributions. One comes from the splitting between the first two generation
scalar masses, ${W_{D_{L(R)}}}_{21} (m_1^2-m_2^2)$. We can use the 
constraints in \cite{HKT,GMS} in this case because the first two 
generation scalar masses have to be degenerate to a high degree and there
is no ambiguity in what $M_{\tilde{q}}$ is. The other comes from the large
splitting of the third generation scalar mass, ${W^*_{D_{L(R)}}}_{32}
{W_{D_{L(R)}}}_{31} (m_1^2-m_3^2)$. This part can be treated in the
same way as in the $B-\bar{B}$ mixing described above.
The terms proportional to the left-right mass insertions are a little
more complicated because they involve new integrals. These terms are
proportional to $[m_b\,(A+\mu \tan\beta)]^2$. For our purpose, we always 
work in the large \tb\ scenario.
Hence the corresponding
constraints scale as ${m_3^3\over \mu\tan\beta}$, versus $m_3$ in the 
case of chirality-conserving terms. The results are listed in Table B1(b)
for $\Delta m_K$ and Table B1(c) for $\epsilon$.
The $\epsilon'$ parameter could put constraints on
$|\Im \wdlr^*_{32} \wdlr_{31}|$ and $|\Im \wdlr_{32} \wdrl_{31}|$.
The first one is weaker than the constraints from other places,
the second one is enhanced by \tb\ and is listed in Table B1(d).
The numbers are obtained by requiring its contribution to
$\epsilon'$ smaller than $3\times 10^{-3} \epsilon$.

The mixing matrix elements ${W_{D_{L(R)}}}_{32}$ are constrained by
the $b\to s\gamma$ decay. The $b\to s\gamma$ branching ratio has been
measured to be $(2.32\pm 0.57 \pm 0.35)\times 10^{-4}$ by CLEO
\cite{CLEO}, which is consistent with the Standard Model prediction
$(2.8\pm 0.8)\times 10^{-4}$\cite{Buras}. In supersymmetric models 
there are many other contributions. The gluino diagram contributions 
depend on the mixing matrix elements ${W_{D_{L(R)}}}_{32}$ so they
can be used to constrain  ${W_{D_{L(R)}}}_{32}$. Unlike other contributions,
the gluino diagrams give significant contributions to both
$\bar{s}_L \sigma^{\mu \nu} b_R F_{\mu \nu}$ and
$\bar{s}_R \sigma^{\mu \nu} b_L F_{\mu \nu}$ operators.
The former can interfere constructively or destructively with other
contributions and the latter does not. In Table B1(e) we list the constraints
on $\wdl_{32}$ and $\wdr_{32}$ by requiring that each gluino diagram alone
does not exceed the Standard Model contribution.

The up mixing matrices $W_U$'s are constrained by $D-\bar{D}$ mixing, and
the results are shown in Table B1(f).

In the lepton sector, the most stringent constraints come from $\mu \to
e\gamma$ decay. In the large \tb\ scenario in which we are interested,
the amplitude of the dominant contribution is given in Ref. \cite{ACH}.
Requiring that the rate does not exceed the experimental limit, 
$B(\mu\to e\gamma) < 4.9\times 10^{-11}$\cite{B} give constraints on
${W_{E_{L(R)}}}_{32} {W_{E_{R(L)}}}_{31}$, which are shown in Table B1(g).
Because we are interested in generating $m_e$ by radiative corrections
which requires sizable mixing between the first and the third generations,
${W_{E_{L(R)}}}_{31}$, the $\tau\to \mu\gamma$ decay does not give stronger
constraints on ${W_{E_{L(R)}}}_{32}$ than those from the 
$\mu\to e\gamma$ decay.

\begin{center}
\vskip .2in
{\bf Table B1}
\nopagebreak
\vskip .1in
(a) {\boldmath$\Delta m_B$}
\nopagebreak

\begin{tabular}{|c|c|c|} \hline
$\sqrt{y}$ & \footnotesize{$\sqrt{|\Re (\wdlr_{31})^2 |}$} &
  \footnotesize{$\sqrt{|\Re (\wdl^*_{31}\wdr_{31})|}$} \\ \hline
$2$ & $1.0\times 10^{-1}$ & $3.1\times 10^{-2}$ \\ \hline
$3$ & $6.5\times 10^{-2}$ & $2.4\times 10^{-2}$ \\ \hline
$5$ & $4.9\times 10^{-2}$ & $2.0\times 10^{-2}$ \\ \hline
\end{tabular}

\vskip .2in
(b) {\boldmath$\Delta m_K$}
\nopagebreak

\begin{tabular}{|c|c|c|c|} \hline
$\sqrt{y}$ & \footnotesize{$\sqrt{|\Re (\wdlr^*_{32}\wdlr_{31})^2 |}$} &
  \footnotesize{$\sqrt{|\Re (\wdl^*_{32}\wdl_{31}\wdr_{32}\wdr^*_{31})|}$} &
  \footnotesize{$\sqrt{|\Re (\wdlr_{32}\wdrl_{31})^2|}^\#$}\\ \hline
$2$ & $4.7\times 10^{-2}$ & $5.6\times 10^{-3}$ & $7.4\times 10^{-2}$ \\ \hline
$3$ & $3.0\times 10^{-2}$ & $4.2\times 10^{-3}$ & $4.7\times 10^{-2}$ \\ \hline
$5$ & $2.2\times 10^{-2}$ & $3.6\times 10^{-3}$ & $3.7\times 10^{-2}$ \\ \hline
\end{tabular}

\vskip .2in
(c) {\boldmath$\epsilon$}
\nopagebreak

\begin{tabular}{|c|c|c|c|} \hline
$\sqrt{y}$ & \footnotesize{$\sqrt{|\Im (\wdlr^*_{32}\wdlr_{31})^2 |}$} &
  \footnotesize{$\sqrt{|\Im (\wdl^*_{32}\wdl_{31}\wdr_{32}\wdr^*_{31})|}$} &
  \footnotesize{$\sqrt{|\Im (\wdlr_{32}\wdrl_{31})^2|}^\#$} \\ \hline
$2$ & $3.7\times 10^{-3}$ & $4.6\times 10^{-4}$ & $6.0\times 10^{-3}$ \\ \hline
$3$ & $2.4\times 10^{-3}$ & $3.4\times 10^{-4}$ & $3.8\times 10^{-3}$ \\ \hline
$5$ & $1.8\times 10^{-3}$ & $2.9\times 10^{-4}$ & $3.0\times 10^{-3}$ \\ \hline
\end{tabular}

\vskip .2in
(d) {\boldmath$\epsilon^{\prime}$}
\nopagebreak

\begin{tabular}{|c|c|} \hline
$\sqrt{y}$ & \footnotesize{$|\Im (\wdlr_{32} \wdrl_{31})|^\# $} \\ \hline
$2$ & $1.4\times 10^{-3}$ \\ \hline
$3$ & $7.7\times 10^{-4}$ \\ \hline
$5$ & $5.4\times 10^{-4}$ \\ \hline
\end{tabular}

\vskip .2in
(e) {\boldmath $b \rightarrow s \gamma$}
\nopagebreak

\begin{tabular}{|c|c|} \hline
$\sqrt{y}$ & \footnotesize{$|\wdlr_{32}|^\# $} \\ \hline
$2$ & $6.9\times 10^{-2}$ \\ \hline
$3$ & $5.3\times 10^{-2}$ \\ \hline
$5$ & $4.7\times 10^{-2}$ \\ \hline
\end{tabular}

\vskip .2in
(f) {\boldmath $\Delta m_D$}
\nopagebreak

\begin{tabular}{|c|c|c|c|} \hline
$\sqrt{y}$ & \footnotesize{$\sqrt{|\Re (\wulr^*_{32}\wulr_{31})^2 |}$} &
  \footnotesize{$\sqrt{|\Re (\wul^*_{32}\wul_{31}\wur_{32}\wur^*_{31})|}$} &
  \footnotesize{$\sqrt{|\Re (\wulr_{32}\wurl_{31})^2|}^\#$} \\ \hline
$2$ & $9.5\times 10^{-2}$ & $3.0\times 10^{-2}$ & $3.9\times 10^{-1}$ \\ \hline
$3$ & $6.3\times 10^{-2}$ & $2.3\times 10^{-2}$ & $2.5\times 10^{-1}$ \\ \hline
$5$ & $4.7\times 10^{-2}$ & $1.9\times 10^{-2}$ & $2.0\times 10^{-1}$ \\ \hline
\end{tabular}

\vskip .2in
(g) {\boldmath $\mu \rightarrow e \gamma$}
\nopagebreak

\begin{tabular}{|c|c|c|} \hline
$\sqrt{y}$ & \footnotesize{$|\welr^*_{32}\welr_{31}|^\#$} &
  \footnotesize{$|\welr_{32}\werl_{31}|^\#$} \\ \hline
$2$ & $2.4\times 10^{-3}$ & $2.2\times 10^{-4}$ \\ \hline
$3$ & $1.8\times 10^{-3}$ & $1.3\times 10^{-4}$ \\ \hline
$5$ & $1.6\times 10^{-3}$ & $1.0\times 10^{-4}$ \\ \hline
\end{tabular}
\end{center}

\vskip .2in
\noindent {\bf Table B1}: 
Constraints on the fermion-sfermion flavor mixing matrix
elements. The reference values are taken as: $\tilde{m}_3 
= M_g =500\, $GeV, $\mu=500\,$GeV, $\tan\beta =60$, and 
$\sqrt{y}\equiv {\tilde{m}_1 \over \tilde{m}_3}$. The ones with \# scale
as $({\tilde{m}_3 \over 500GeV})^3 ({500GeV \over \mu})
({60 \over \tan\beta})$, others scale as ${\tilde{m}_3 \over 500GeV}$.


\newpage
\begin{center}
{\bf Figure Captions}
\end{center}

\noindent Fig.\ 1: The dominant radiative contributions to the fermion masses:
  (a) charged leptons, (b) up-type quarks, (c) down-type quarks.
\vskip .2in

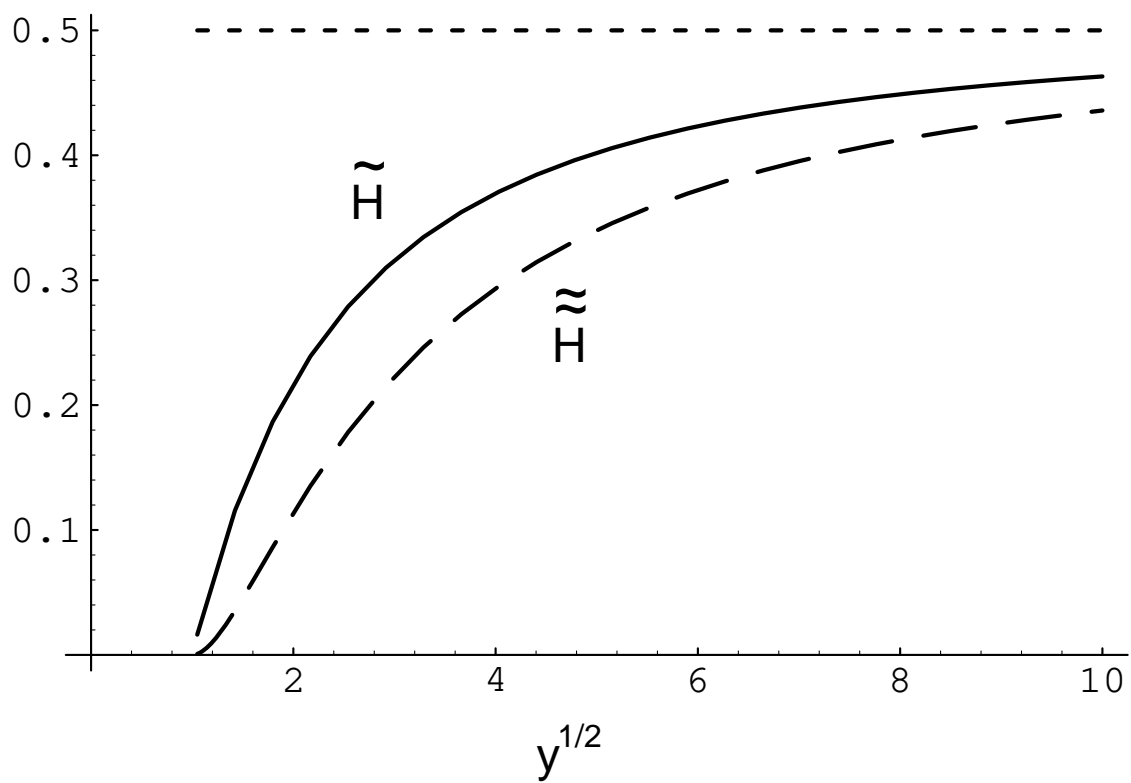
\noindent Fig.\ 2: Plots of the super-GIM factor 
  $\wt{\wt{H}}\equiv h(x_3, x_3)
  -h(x_3, x_1) - h(x_1, x_3) + h(x_1, x_1)$ and $\wt{H} \equiv h(x_3, x_3)
  -h(x_3, x_1)$ versus the ratio between
  the first two generation and the third generation scalar masses
  $\sqrt{y}$, $y\equiv \tilde{m_1}^2/\tilde{m_3}^2 = x_1/x_3$, with 
  $x_3=1,\, (M_g=\tilde{m_3})$. 
  \mbox{${\Delta {\bf m}_{e\alpha\beta}\over m_\tau}
  = 2.4\times 10^{-2} ({\mu\over m_{\tilde{\tau}}})({\tan\beta\over 60})
  ({\wt{\wt{H}}\over 0.5}) \sqrt{x_3}\wel_{3\alpha} \wer_{3\beta}$,} \\
  \mbox{${\Delta {\bf m}_{u\alpha\beta}\over m_t}
  = 1.2\times 10^{-2} ({A\over m_{\tilde{t}}})
  ({\wt{\wt{H}}\over 0.5}) \sqrt{x_3}\wul_{3\alpha} \wur_{3\beta}$,} \\
  \mbox{${\Delta {\bf m}_{d\alpha\beta}\over m_b}
  =0.7 ({\mu\over m_{\tilde{b}}})({\tan\beta\over 60})
  ({\wt{\wt{H}}\over 0.5}) \sqrt{x_3}\wdl_{3\alpha} \wdr_{3\beta}$,} \\
  for $\alpha,\,\beta = 1,\,2$, and $\wt{\wt{H}}$ has to be replaced
  by $\wt{H}$ if one of the $\alpha,\,\beta$ is 3.
\vskip .2in

\noindent Fig.\ 3: Contour plot of $B(\tau\to e\gamma)$, where the
mixing angles are fixed by requiring a radiative electron mass. We have
put $\tan \beta$ = 60., $\mu$ = $m_{\tilde{\tau}}$=200 GeV, 
and plot in the $M_1$ -- 
$\sqrt{y}$ plane where $M_1$ is the bino mass and we have assumed 
the GUT relation $M_2 \sim 2 M_1$; 
$y ={{m^2_{\tilde{e}}}\over{m^2_{\tilde{\tau}}}}$. We also assume that the
left and right handed mixing angles are equal, giving us a lower bound on
$B(\tau\to e\gamma)$.
The branching ratio scales as ${{\mu \tan \beta}\over {m^4_{\tilde{\tau}}}}$.
\vskip .2in

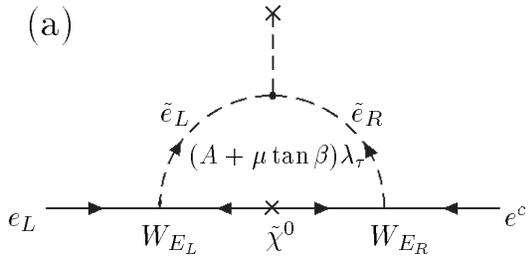
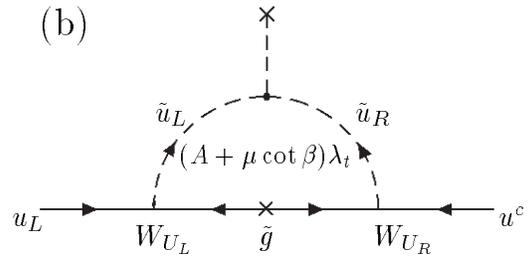
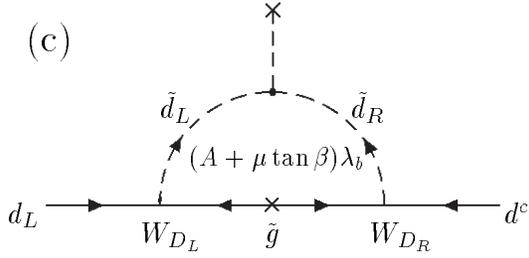
\noindent Fig.\ 4: Chargino diagrams which contribute to radiative down-type
  quark masses and are enhanced by large $\tan\beta$. 
\vskip .2in

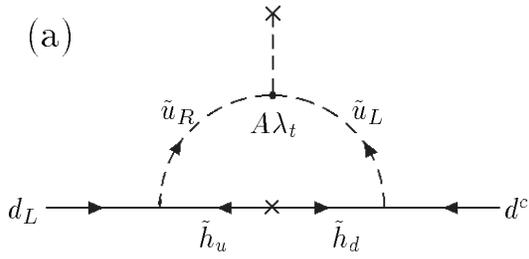
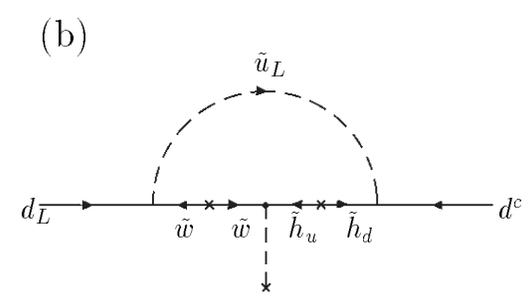
\noindent Fig.\ 5: The diagram which generates the second generation masses.
\vskip .2in

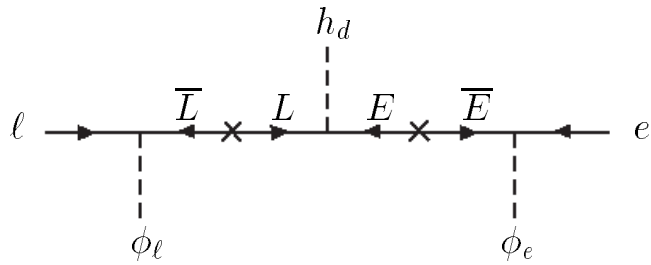
\noindent Fig.\ 6: $D$ term mixing between the first and the third generations.

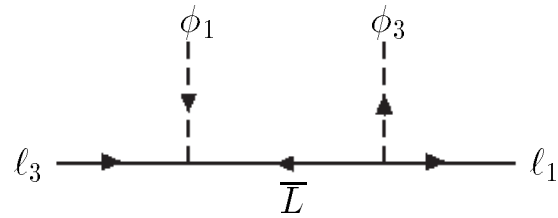
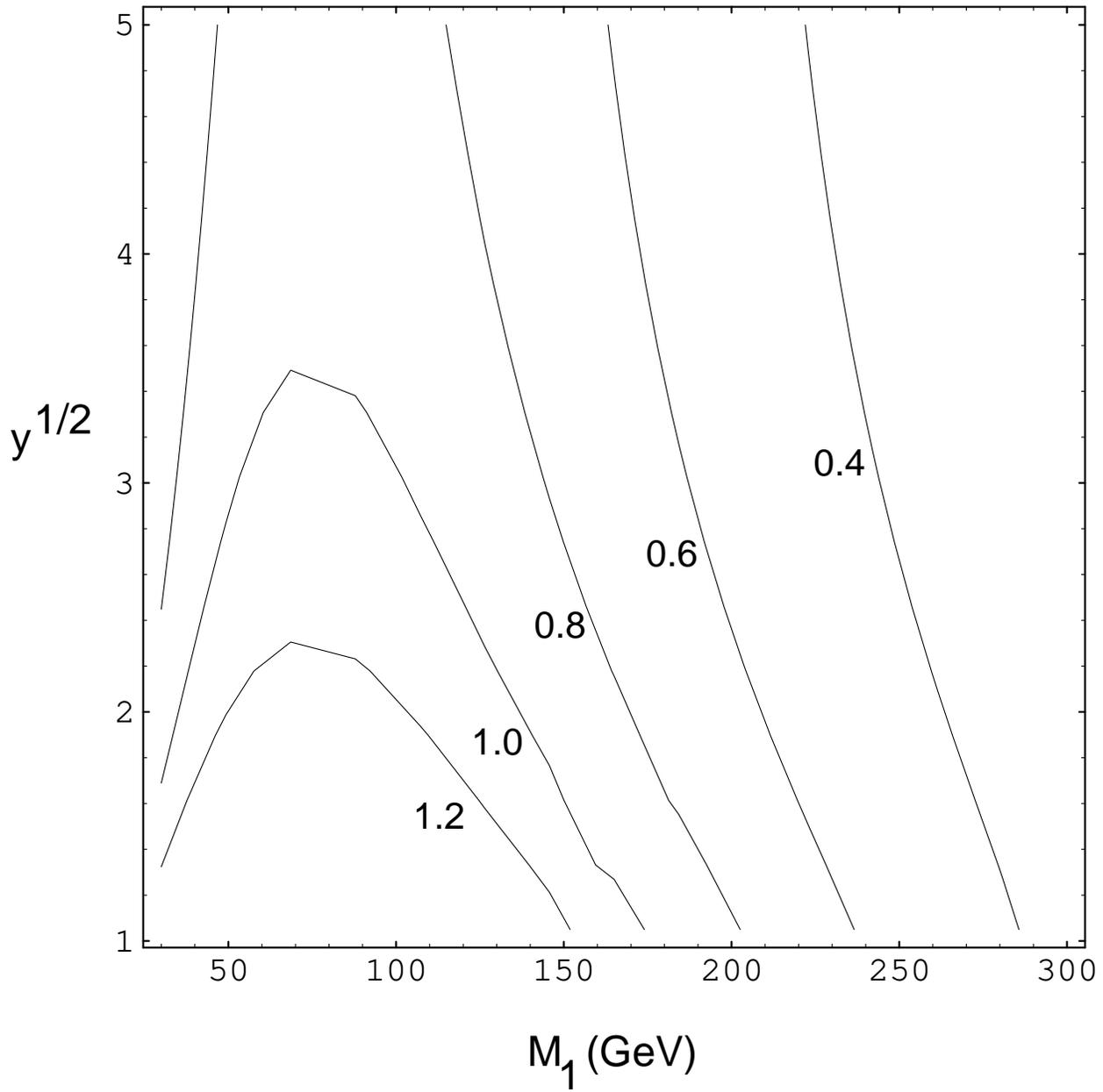
\end{document}